# Meraculous2: fast accurate short-read assembly of large polymorphic genomes

Jarrod A. Chapman[1], Isaac Y. Ho[1], Eugene Goltsman[1] Daniel S. Rokhsar[1,2]

[1] Department of Energy Joint Genome Institute, 2800 Mitchell Drive, Walnut Creek, CA 94598

[2] Department of Molecular and Cell Biology, University of California, Berkeley, CA 94720

**ABSTRACT**

We present Meraculous2, an update to the Meraculous short-read assembler that includes (1) handling of allelic variation using "bubble" structures within the de Bruijn graph, (2) improved gap closing, and (3) an improved scaffolding algorithm that produces more complete assemblies without compromising scaffolding accuracy. The speed and bandwidth efficiency of the new parallel implementation have also been substantially improved, allowing the assembly of a human genome to be accomplished in 24 hours on the JGI/NERSC Genepool system. To highlight the features of Meraculous2 we present here the assembly of the diploid human genome NA12878, and compare it with previously published assemblies of the same data using other algorithms. The Meraculous2 assemblies are shown to have better completeness, contiguity, and accuracy than other published assemblies for these data. Practical considerations including pre-assembly analyses of polymorphism and repetitiveness are described.

**INTRODUCTION**

Deep and accurate short-read shotgun coverage of human and other gigabase-scale genomes is now readily accessible at modest cost. With these increases in sequencing throughput, a new generation of computational algorithms has been developed to assemble shotgun sequence for large and complex genomes (reviewed in [1-3]). These approaches typically incorporate de Bruijn graph approaches pioneered in short-read assembly by Euler[4] and Velvet.[5] Several groups have assembled human and other mammalian genomes[6-11] with these methods. Such assemblies are challenging not only because of genome size and repetitiveness, but also because of the intrinsic heterozygosity of outbred species, which encompasses single nucleotide variants (SNVs), small and large insertions and deletions (INDELs), and larger structural variation, as have been well characterized for humans.[12]

Assembling large, repetitive, and polymorphic genomes typically requires large shared-memory systems, and runs can take a week or more to produce a single assembly, consuming significant computing resources. While several large genomes have been assembled purely from short (<150 bp) paired reads, it remains unclear which genomes have a structure that will permit a good quality assembly, and/or which combinations of data will facilitate this. Indeed, it is even unclear how to measure assembly quality.[11, 13, 14] The "assemblathon" competitions are useful to serve as a testing ground for new approaches,[11, 13] providing common datasets to facilitate direct comparisons of algorithms and implementations. With large projects like Genome 10K to assemble 10,000 vertebrate genomes[15] the development of efficient and accurate methods to produce complete and high quality assemblies is essential.

Previously, we described the "Meraculous" algorithm for shotgun assembly as a hybrid k-mer/read-based assembler.[16]  Briefly, Meraculous first assembles unique regions of the genome into preliminary uncontested "UU" contigs by efficiently constructing and traversing a simplified de Bruijn k-mer graph. These contigs are then linked together by aligning them to paired-end read data, and gaps in the resulting scaffolds are filled using localized assemblies of relevant reads.  Meraculous capitalizes on the high accuracy of current Illumina sequence by eschewing an explicit error correction step, which we argue is redundant with the assembly process.  In its initial version, Meraculous did not accommodate polymorphic diploid genomes, and although parallelized we reported only a ~15 million base pair haploid fungal genome, along with simulated datasets, and did not report assemblies using real data for large (*i.e.*, gigabase scale) genomes.

Here we describe improvements to Meraculous that overcome these previous limitations, extending changes that were incorporated into the Meraculous Assemblathon II entry.[11]  New features include (1) explicit handling of allelic polymorphism using linear chains of "bubble" structures within the de Bruijn graph, (2) improved gap closing, based on case studies, and (3) an improved scaffolding algorithm that produces more complete assemblies without compromising on scaffolding accuracy.  The speed and bandwidth efficiency of the new parallel implementation has been substantially improved, allowing the assembly of a human genome to be accomplished in 24 hours of real time on the JGI Genepool cluster (see **Table 1** for details of the resources used to perform this computation).

To explore these features of Meraculous2 we describe here the assembly and analysis of human short-read datasets for identifier NA12878, a woman of European ancestry whose genome has been sequenced and extensively analyzed by the 1000 Genomes Project. [12]  By trio phasing (*i.e.*, combining the sequence of NA12878 and her parents), phased maternal and paternal haplotype sequences have been inferred,[17] providing a natural external reference against which shotgun assemblies can be compared.  The genome of NA12878 has been assembled previously by ALLPATHS-LG [9] and SGA,[10] allowing us to compare and contrast the performance of Meraculous2 against these two state-of-the art assemblers. We present several previously undescribed methods for analyzing short-read datasets in the context of this human assembly, and discuss several metrics for measuring the correctness of long-range linkages found in human assemblies.  Prospects for further improvement are discussed.

**METHODS**

**Test datasets.** We downloaded two available datasets for an individual from the CEU HapMap population (identifier NA12878) to allow direct comparisons with previous assemblies for this individual. To compare with SGA we used the same dataset as Simpson and Durbin[10] by downloading 2.5 billion 101 bp reads sequenced by the Broad [9] Institute as part of the 1000 genomes project[12] (ftp://ftp.1000genomes.ebi.ac.uk/vol1/ftp/technical/working/20101201_cg_NA12878/NA12878.hiseq.wgs.bwa.raw.bam). These reads are from a paired-end library of fragments with average size 380 bp, and represent a total of 84-fold redundant coverage. We also downloaded the dataset from Gnerre *et al.* to compare with AllPaths-LG.[9]

**Reference human data**. As a reference sequence for NA12878, we used the December 16, 2012 version (available from http://alleleseq.gersteinlab.org/downloads.html) .[17]

**Fosmids.** The Broad Institute Genomics Platform selected 103 randomly selected fosmids from NA12878 for complete sequencing. B13. These were accessed from Genbank BioProject Accession: PRJNA196715.

**Efficient k-mer counting, UFX graph construction, and UUcontig generation.** Meraculous begins by counting all k-mers for a given fixed odd value of k. For each k-mer that occurs more than $D_{min}$ times, we count the number of "high quality" single base extensions of each type (ACTG) in the read set, where high quality is defined as greater than or equal to a threshold phred quality score $Q_{min}$. Typically we choose $Q_{min}$ = 20. A k-mer that has a unique high quality extension at an end is designated as "U" at that end; k-mers that have multiple supported extensions are marked as "F" (forked), and k-mers with no extensions are "X."  The resulting hash represents the UFX graph.  As in the original Meraculous, this calculation is parallelized by partitioning k-mers according to their prefix in a load-balanced fashion. Each node reads the full input dataset and tallies the mer-counts for the prefixes it is responsible for. This calculation is multithreaded on a node, but without inter-node communication.  "UU" contigs are obtained by traversing the subgraph of k-mers that are designated "U" at each end.  These are unique, unforked paths in the de Bruijn graph.

**Multithreaded approach**.  A common multi-threaded design was applied to the mercounting and mergraph (UU contig generation) modules.   An array of 65,536 Google sparse_hash tables, each guarded by a mutex, is used to store mers as compressed objects that are quickly convertible between text and binary space using lookup tables.

The contig generation module first hashes the UFX graph.   Keys are farmed out across multiple threads and traversed using the UFX codes to generate contigs, one traversal per thread.  A secondary hash is kept for each thread to track the mers that have been visited during the current traversal.  When the traversal is done, the lexicographically least mer from the walk is looked up in a results hash.  If that mer is not found, then the mer is added, the contig is printed, and all of the secondary hash keys for that thread are marked as visited in the primary hash; subsequent traversals will then stop upon visiting these mers.   If the mer is found, then the walk is discarded.

Locking is only needed while looking up and adding the lexicographically least mer in the results hash.

**Bubbletigs.** For diploid genomes, "bubbles" are defined as pairs of UU-contigs that share a common unique k-mer extension at both ends (**Figure 1**). "Diplotigs" are constructed by connecting bubbles and non-bubble UU contigs that terminate in the same unique k-mer extension as the bubble into alternating chains of the form contig-(bubble-contig)$^n$. The depth of a diplotig is reported as the base-weighted average of the depth of its constituent UU contigs. Isolated UU contigs (of length at least twice the mer-size) are reported as "isotigs" to be used in scaffolding; typically only isotigs with depth consistent with homozygosity are retained for this purpose. The collection of diplotigs and isotigs are collectively referred to as "bubbletigs." By default, the sequence reported for a diplotig in its bubble regions is that of the maximum depth branch of each bubble.

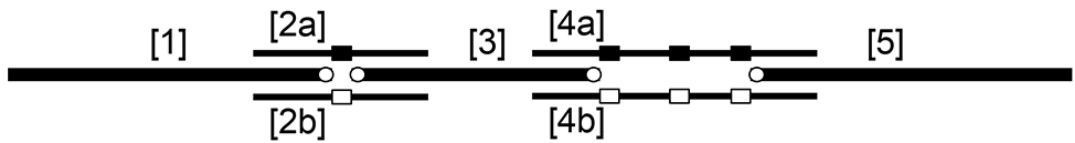

**Figure 1. Bubbles and diplotigs.** UUcontigs are indicated by horizontal lines; open and filled rectangles representing the two alleles at a heterozygous site. UUcontigs [1], [3], and [5] do not include any heterozygous sites. The UUcontig pair [2a]/[2b] represents the two k-mer paths at alternate alleles for an isolated heterozygous site. The UUcontig pair [4a]/[4b] spans multiple heterozygous sites that are closer than k nucleotides apart. Note that UUcontig [1] has two rightward extensions, [2a] and [2b]; this fork is the reason that UUcontig [1] terminates. Conversely, the leftmost k-1 nucleotides of UUcontigs [2a] and [2b] are identical, and occur uniquely as the k-1 nucleotide suffix of a single k-mer that starts with final k-1 nucleotides of [1].

This reciprocal relationship between [1] and [2a]/[2b] is repeated at all "forks," defining a unique bubbled path (a "diplotig") through this region of the genome.

**Mer-mapping algorithm**. To organize UU contigs or bubbletigs (generally, contigs) into scaffolds, we first align all reads to the contig sequences. Since by construction, k-mers occur at most once in the UU contigs or bubbletigs, we can guarantee that alignments represent the exact placement of the read on the assembly by requiring a minimum exact match of k.

The merBlast module replaces the BLAST-based method used previously for aligning reads to contigs.[16] The contigs are hashed in memory as compressed k-mers. Each read is scanned to find the leftmost and rightmost mers that have a hit to the hash; their implied alignments are extended to the end of the read when possible, allowing for at most one indel or mismatch off each end. If the two alignments point to the same region of the genome, then the read is done processing and a single combined alignment is reported. Otherwise, the two alignments are extended inwards by scanning along the read and reported separately; the rest of the read is scanned, and all remaining implied alignments are reported with no allowances for mismatches or indels.

**Scaffolding algorithm.** Scaffolding is typically performed as a hierarchical process, iterated over libraries of increasing length scale. Read pair placements from related libraries are analyzed to construct "links" between sequence objects (either between contigs directly at the base of the scaffolding hierarchy, or between scaffolds from a previous iteration).  Links may be determined using individual libraries, or groups of similarly-sized libraries.  Links represent the number of pairs that connect two sequences and may be either "splinting" (a single read aligning across two sequences) or "spanning" (a pair of reads aligning to two separate sequences).

The resulting scaffold graph represents each sequence-end as a vertex and each link between sequences as an edge.  To produce scaffolds this graph is traversed as follows:

1. Sequence-ends are first checked for topological defects in linkage (links to themselves, or the opposite end of the same sequence; multiple conflicting links; links to sequences with a significantly different depth estimate than themselves).  Ends without a defect are allowed to participate in the traversal.
2. "Long" sequences (*i.e.,* those whose length is greater than half of the size of the largest insert size used in the current round of scaffolding) are connected to their unique closest long eligible sequence, *i.e.*, the long sequence with the shortest estimated separation from the sequence under consideration.
3. If no such long sequence exists, sequences are connected to their closest available "extendable" (*i.e.*, having further downstream links) linked sequence.
4. If no extendable sequences are linked, the closest available linked sequence is used.  Such non-extendable links terminate the scaffold.
5. Sequences that remain unconnected may be suspended between pairs of linked sequences if they link uniquely between sequences connected previously in the traversal.

A scaffold-report-format ("srf") file is produced representing the order and orientation of the original UU contigs (or bubbletigs), their estimated depth, and the estimated size of each inter-contig gap.  The srf-file may be used to produce scaffold fasta directly (with or without additional gap-closing) or as input to additional rounds of scaffolding using longer-range libraries.  In the latter case, read alignments are projected forward from the original contigs to the current scaffolds and the process iterated.  Because the scaffold-generation step is fast and inexpensive, explicit optimization over the minimum number of read pairs in a traversed link is performed at this step, choosing the scaffolding which maximizes scaffold N50 at each level of the scaffolding hierarchy.

**Gap-closing.**  After constructing UU contigs (or bubbletigs in a diploid case), and forming scaffolds by mapping paired-end reads, we attempt to close remaining gaps that are "captured" within scaffolds using reads that (a) extend into a gap based on their (partial) alignment to one or both flanking contigs, and (b) unaligned reads that are inferred to lie in or near the gap based on the aligned position of their paired-end read. Unaligned reads are assigned to each of those gaps whose footprint overlaps the position of the read as estimated from the aligned position of its paired-end (using the mean insert size of the fragment library plus and minus two standard deviations to perform this placement estimate).

Projected reads are used to close gaps as follows:

1. Identify "splint" reads that are anchored by k-mers found on both sides of the

gap. If at least two splinting reads are found and all such reads agree on the gap sequence, the gap is resolved. If no such splints are found:

2. Seek a unique "right-walk" using k-mer extension from the left boundary of the gap to the right boundary of the gap. In these walks, only k-mers present in the reads that are projected into a gap are used. If no right walk is found, a "left-walk" is sought using k-mer extension from the right boundary to the left.

3. "Patching" together the maximal right-walk and maximal left-walk, requiring an overlap of at least 10 bases. (If no such patch is detected, these maximal walks are discarded, and the gap is left unclosed.)

Any gap resolution suggested by (1)-(3) must have a length that is within three standard deviations of the estimated gap size, based on linking information. Gap closures outside of this range are not accepted by default, but may be accepted if the user requests "aggressive" gap closure.

The "mer-walking" step (2 above) uses a modified version of the contig generation extension algorithm[16] applied to the projected reads. In this modification:

1. Extension bases are categorized by phred quality score into four bins (0-10, 11-20, 21-30, >30) and extensions are accepted or rejected based on the number and quality category of extending bases. This allows for uncontested extensions of lower quality than used in the original contig generation to be accepted in gap-closing.

2. The mer-size used is dynamically adjusted by "upshifting" (mer-size increased by 2) when a fork is encountered, or "downshifting" (mer-size decreased by 2) when a termination is encountered. Repeated upshifting or downshifting will occur until the gap is resolved or the data is no longer sufficient to support extension.

Polymorphic gap-closing is used in the case of diploid genomes by removing the uniqueness criterion of the "splinting" method by allowing the maximum-frequency "splint" sequence to resolve the gap (if one exists); or failing that, allowing a single maximum-frequency choice at a forking position in the "mer-walking" method.

Gaps in repetitive scaffolds (i.e., high depth scaffolds, as defined by a repeatCopyCount parameter, typically set to 2 to identify scaffolds with double-depth or higher) may be optionally excluded from gap-closure as these represent multiple loci in the genome and any attempted resolution could result in effectively chimerizing the disparate repeat copies.

Gap-closing is parallelized by decoupling the projection of reads into gaps from the computation of the gap resolution. In this way each gap can be closed independently and the closure calculations may be distributed among an arbitrary number of compute nodes. The gap-resolutions are then combined with the scaffold report and the contig sequence to produce a final gap-closed consensus sequence. In the generation of the final consensus, unresolved gaps are represented by a string of Ns of length equal to the estimated size of the gap. By convention, if the estimated gap size is smaller than a user-specifiable parameter (minGapNs), the flanking contigs are trimmed back to create a gap no smaller than this minimum. This process eliminates potential redundancies in the contig sequences flanking the gap and avoids coordinate drift due to consistent over representations of gap size due to the use of a minimum run of N's.

## RESULTS

To demonstrate Meraculous2 we assembled a combination of well-studied shotgun datasets that have been used as benchmarks in previous studies. We used shotgun data for individual NA12878, a woman of European ancestry whose genome was analyzed by the 1000 Genomes pilot,[12] assessed for structural variants[18], and haplotyped[19]. We walk through the process of assembly, describing our approach including diagnostics and heuristics that were applied.

**Mercounting, heterozygosity, and genome structure**. The first step in a Meraculous assembly is to count the number of times each k-mer occurs in the dataset, for selected (odd) k. Counting is parallelized across many nodes by assigning each processor its own load-balanced set of k-mer prefixes, as described previously[16] (Methods). Only k-mers that occur at least $D_{min}$ times in the data are retained in subsequent steps, with $D_{min}$ typically 2 or 3. This substantially reduces the number of distinct k-mers to consider, since most unique or low-count k-mers arise from errors. Typically we only mer-count fragment libraries, since mate-pair libraries may have coverage non-uniformities, redundancies, and/or undetected chimerism.

The distribution of k-mer frequencies is expected to follow the Poisson distribution for random sampling of a sequence, but is typically broader due to non-uniformities in sequencing (**Figure 2a**). For frequencies near the nominal "full" k-mer depth $d_0 = N(L-k+1)/G$ (where N is the total number of reads, L is the read length, and G is the genome size), the distribution is often well-fit by a normal distribution for a haploid or homozygous genome. The peak frequency is typically reduced from its nominal theoretical value due to sequencing errors. For a heterozygous genome, a second peak at half depth appears, representing k-mers that span allelic variants (**Figure 2b**). Relative to the homozygous case, for each isolated heterozygous single nucleotide variant, the full depth peak is depleted by k k-mers, and the half-depth peak is augmented by 2k k-mers. This model can be used to estimate the heterozygosity of a genome prior to assembly by fitting the k-mer frequency distribution to a sum of two normal distributions with mean $d_{max}$ and ½ $d_{max}$ (**Supplementary Note 1**).

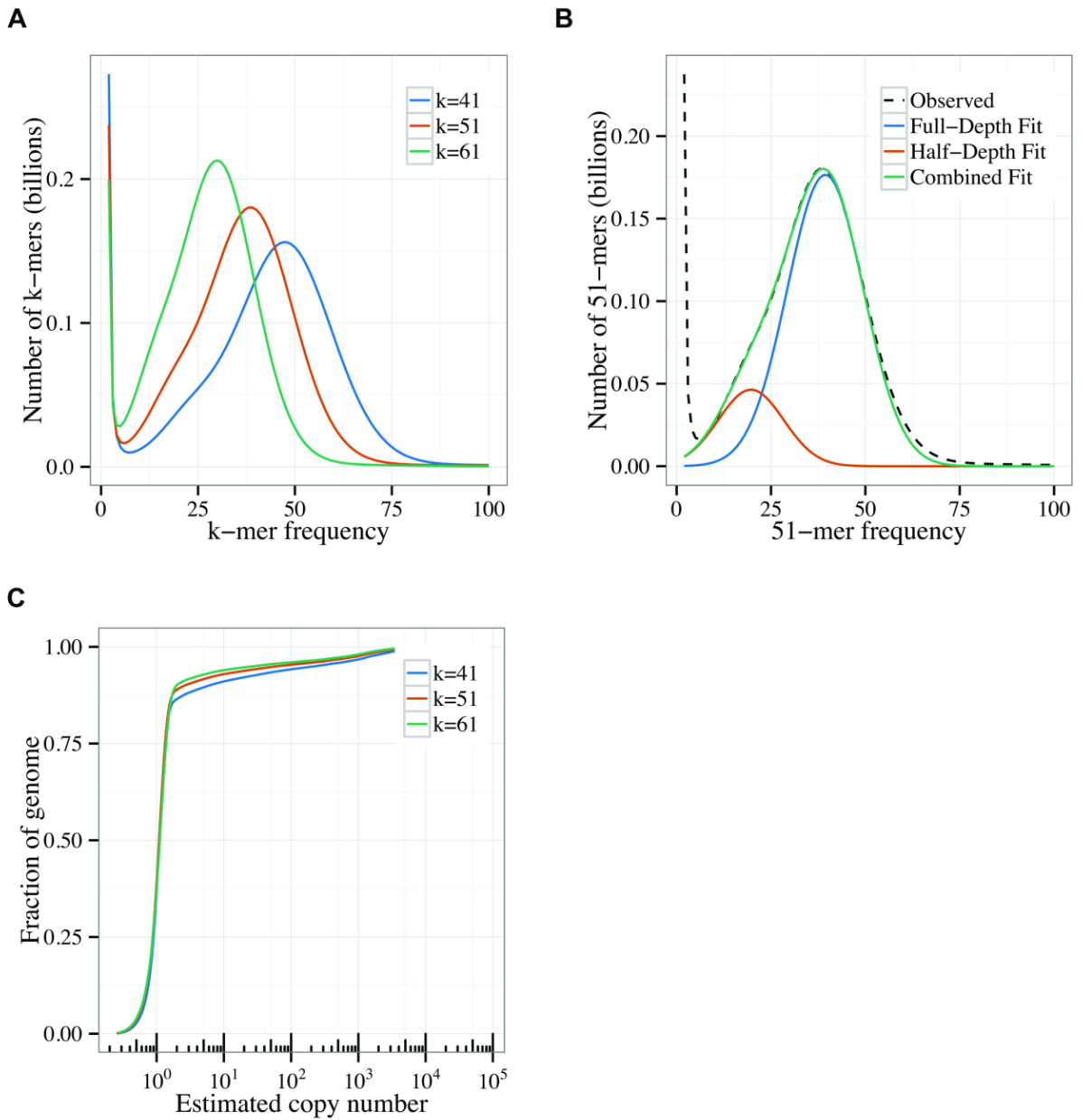

**Figure 2. k-mer distributions for the human NA12878 short-fragment dataset.**

Panel (A) shows the k-mer spectra for k=41, 51, and 61. As *k* increases at fixed read size *L*, the k-mer frequency shifts to lower depth since each read contains *L-k+1* k-mers. (B) Skewed k-mer frequency distributions include a full-depth peak representing homozygous loci and a half-depth peak representing heterozygous sites. Fitting with two normal distributions allows heterozygosity to be estimated. Note that each polymorphic site contributes *2k* k-mers to the half-depth peak.

(C) Cumulative distribution for k=41, 51, and 61.  Here the horizontal axis, shown on a logarithmic scale, is measured relative to the peak depth, assumed to represent single copy sites in the genome.  This cumulative distribution allows an estimate of genome size and the fraction of the genome that is single copy for a given k.

While the k-mer frequency distribution in the vicinity of $d_0$ is determined by heterozygosity, the full distribution across all depths measures the fraction of the genome that occurs in different copy number on the scale of a k-mer. A simple but useful characterization of the size and assembly complexity of a genome is then given by the cumulative number of distinct k-mers that occur in the dataset with frequency less than or equal to x.  It is convenient to exclude k-mers that occur fewer than ¼ $d_{max}$ times in the data (since these often arise due to errors), and to plot the cumulative distribution versus the logarithm of scaled depth $x/d_o$ (**Figure 2c**).  For the human genome, and indeed all genomes that we have studied in this manner, this cumulative distribution has a sharp knee that is characteristic of the genome and that varies with k. This knee separates the fraction of the genome that is in single copy sequence (with respect to the specified k) and is therefore relatively easy to assemble from high copy sequences that are difficult to assemble.  Plotting this cumulative distribution vs copy number on a logarithmic scale emphasizes the very rapid growth of repeat copy number -- most sequences are either single copy, or high copy (greater than 10 or 100).

These analyses of the k-mer frequency distribution (**Figure 2**) provide assembly-free estimates of (a) the total size of the genome, (b) the fraction of the genome that can be expected to be readily assembled into contigs, and (c) heterozygosity.  Depending on the detailed structure of the genome, additional multiple-copy sequences can also be assembled if they are interspersed among single copy sequence on the scale of a read length and/or insert length. Our rule of thumb is that we choose k as large as feasible to maintain peak depth ~30 or more. For human we report here k=51. Relative to other genomes, the human genome is relatively repeat-poor at this k-mer length scale, with its knee at ~90%.

**Contigs, bubbles, and diplotigs**. Given a collection of k-mers that occur at frequency $D_{min}$ or greater, Meraculous annotates every k-mer with a code that indicates its status within the Meraculous "UFX" de Bruijn graph. For each occurence of a k-mer in the shotgun read dataset, we catalog the nucleotides that precede and follow it (including occurrences on both strands). Only "high quality" extensions are considered, *i.e.*, those with base quality $Q \geq Q_{min}$, where $Q_{min}$ is typically set to 20 (nominal 1% error rate). These preceding and following nucleotides are referred to as high quality extensions. If a k-mer end has a unique high quality extension (that is, is always followed by the same nucleotide in all high quality instances in the read set) then that end of the k-mer is marked as "U" for unique. If there are multiple high quality extensions, that end of the k-mer is marked as "F" for forked. If there are no high quality extensions (*e.g.*, each occurrence of the k-mer is at the end of a read, or followed by a low quality call), that end of the k-mer is marked with an "X."

A "UUcontig" is a path through the de Bruijn graph such that (a) each k-mer in the path is "UU" (i.e., "U" at both ends), and (b) the steps in the path are reciprocally unique. The reciprocal uniqueness condition controls for forked paths in which one path leads immediately to a dead-end (sometimes called "hair".) [5]  After constructing the UFX graph, the next step in generating an assembly is to construct all UUcontigs, which can be done efficiently in a multithreaded manner (Methods).

UUcontigs represent paths within the de Bruijn graph that are uncontested by any high quality read data, and are the units from which the full contigs and scaffolds are constructed.  The distribution of UUcontig lengths for human is shown in **Figure 3a**.  The sharp peak at contig length 2k-1 corresponds to pairs of short UUcontigs that cover the two alternate alleles of an isolated single nucleotide variant site.  The single polymorphic site lies in the middle of these short UUcontigs (**Figure 1**), and each biallelic site generates a pair of short contigs each representing one allele.  Other heterozygous sites (*e.g.*, indels, complex substitutions) also appear as pairs of short UUcontigs that represent alleles (or, for longer features, haplotypes).  A hallmark of these pairs of short polymorphism-induced contigs is that their preceding (and following) single nucleotide extension kmers are shared, and point uniquely to a single UUcontig in each direction (**Figure 1**).   We refer to such matched pairs of UUcontigs as "bubbles."  The vast majority of bubbles correspond to single nucleotide variants, but indels and other variants are found, as shown by the distribution of length differences between the paired UUcontigs in a bubble (**Figure 3b**).  As expected for heterozygous regions, most paired UUcontigs in bubbles are at half depth (**Figure 3c**). A minority, however, have full depth, which indicates that they represent fixed differences between two-copy repetitive sequences.  Other configurations are observed.

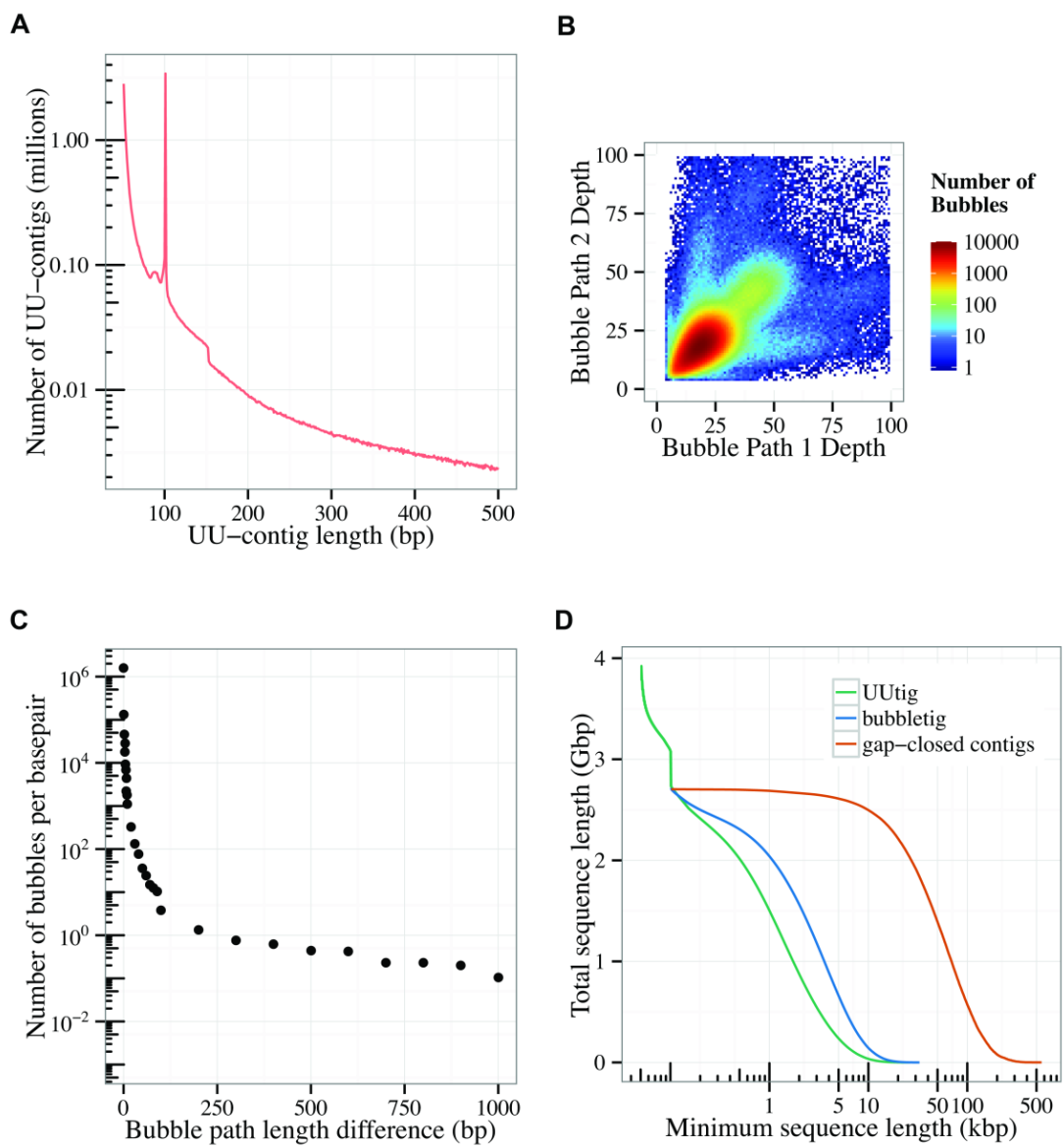

**Figure 3. Features of UUcontigs and bubbletigs.** Panel (A) shows the distribution of UUcontig lengths for the human NA12878 short read dataset. Spikes in the distribution at length 2k-1 correspond to the pairs of such UUcontigs generated by each isolated heterozygous single nucleotide site. (See, *e.g.*, Figure 1). (B) Distribution of path lengths between the paired UU contigs of a bubble, showing that bubbles can represent insertion/deletion variants of hundreds or more nucleotides. (C) heatmap of the number of bubbles vs. the k-mer depth of the two UUcontig pairs in a bubble. (D) Total amount of sequence in UUcontigs and diplotigs longer than a specified length on the horizontal axis. This length is shown on a logarithmic scale.

The philosophy of Meraculous is that in constructing contigs, only uncontested k-mer walks should be used.  While this simple prescription works well for haploid genomes, we can also incorporate heterozygosity in a diploid genome by noting that a bubble represents a unique linkage across a polymorphic site.  So we can extend the UUtig approach to include heterozygosity by allowing linear chains of the form UUtig-bubble-UUtig-bubble…-bubble-UUtig.  We refer to these chains as "diplotigs," which represent uncontested, heterozygosity-containing stretches of the genome.  The cumulative distribution of diplotig lengths longer than $x$ is shown in **Figure 3d**.  Comparison with the UUtig lengths shown in **Figure 3a** show that including bubbles allows much longer chains of sequence to be reconstructed from the de Bruijn graph (N50 improving from 0.5 to 2.5 kbp).

The most extreme allelic divergence observed in human by this method is a heterozygous bubble from the MHC locus. (**Figure 4**).  The two allelic UUcontigs are 295 bp and 299 bp long, respectively; by construction, the first and last 50 bp of these two UUcontigs match.  In the intervening ~200 bp, there are 10 single nucleotide substitutions and one 4 bp gap that accounts for the difference in allelic lengths.  This locus therefore shows a remarkable ~5% divergence between the two NA12878 haplotypes, but is nevertheless represented in our assembly as a single locus, rather than two split contigs, based on its topological organization within a diplotig.

```
        ttgagaagaatgtgtattatatctaaaacatgctaaaaaaattcaccaaa🔵gtgctattag 60
                                                           acgctattag 60

 61  aa----taaatgaatacagtaaagttgcagcaaataaaatcaacacacaaaaatcagtag 116
 61  aactaataaatgaatacagtaaagttgcagcatataaaatcaacacacaaaaatcagtag 120

117  tatttctatatgctaacaatgaactattcaaaaaagaaatcaagaaaacaatcccattta 176
121  tatttctatacactaacaatgaactattcaaaaaacaaatcaagaaaacaatcccattta 180

177  caatatctacaaacaatataaaattcacatggaaccacacaaaaagcctgaatagctaag 236
181  caatatctatgaacaatataaaattcatatggaaccacacaaaaagcctgaatagctaag 240

237  acaatctta
                 🔵agcaaaaagagcaaaactggagacattacaccacttgacttcaaactata
241  acaatcttg
```

**Figure 4. The most divergent bubble in the human NA12878.** This most divergent bubble corresponds to a polymorphic position in the MHC region. Blue circles indicate the forked positions in the k-mer graph that begin and end the bubble. Red sequence indicates polymorphic positions.

In the current implementation, the sequence of a diplotig is represented by the most frequently sampled allele at each bubble. (Note that for a heterozygous genome, the two alleles have equal frequency p=q=1/2, but the read count of each allele will not be equal in a shotgun dataset due to sampling fluctuations.) Using the most frequent allele in consensus produces a mosaic of the two input haplotypes that retains some local phasing information, since we expect fluctuations in the read depth of each allele to be correlated on the scale of a read pair. Information about alternate alleles is provided.

**Scaffolding and gap closure**. The initial stages of Meraculous focus on the k-mer content of reads but do not make use of the inherent contiguity of read sequences (which contain L-k+1 overlapping k-mers) or their pairing relationships. The final stages of the algorithm reintroduce this information by mapping reads to their unique placement relative to contigs.  This allows us to (1) produce high confidence scaffolds, and (2) use read sequences that extend into or are inferred to lie within intra-scaffold gaps to close those gaps between contigs.  We note that by construction k-mers that occur within contigs occur exactly once in the set of bubbletigs (*i.e.*, diplotigs and isolated UU-contigs). To map each read to its corresponding contigs it therefore suffices to find the unique occurrence of its k-mers in the contig set.

To produce scaffolds, Meraculous uses a hierarchical approach (Methods), again following the principle that uncontested linkages should be used at all scales.  Reads are mapped back to the UUtigs (or bubbletigs for a heterozygous genome) taking advantage of the property that each k-mer in the UUtig set occurs once and only once. Linkages are incorporated working iteratively from the shortest to the longest length scales.  At the shortest scale, we identify reads that align at the ends of two UUtigs. These "splints" provide evidence that (1) the two UUtigs should be linked, and (2) what sequence should be used to bridge the gap.  In some cases, splints can be negative, if the two UUtigs overlap with each other on the genome but cannot be joined by a high confidence k-mer walk.   "Spans" are read-pairs such that each read aligns to a different contig.  As described in Methods, the graph of all contig-contig linkages is traversed to identify uncontested paths, and this process is iterated over libraries of increasing insert size.

Once scaffolding is complete, for each intra-scaffold gap we collect reads that either extend into, or are inferred to lie within an intra-scaffold gap. Our aim is then to use these reads to fill gaps by connecting the flanking UUcontigs at either side of a gap by a continuous sequence path. Since reads are anchored relative to the gap based on their own alignment to nearby contigs, or by the alignment of their paired-end, their strand relative to the gap is known. The flanking k-mers of each gap are unique in the UUcontigs, and serve as anchors for seeking such a path. Based on paired-ends and mate-pair data, we have an estimate of the gap size that takes into account ascertainment biases (**Figure 5**).[16] (A new gap size estimator is described in **Supplementary Note 2**.) In some cases gaps have a negative estimated size; this occurs when the left and right flanking contigs overlap, but do not share a path of UU k-mers relative to the entire dataset. In these cases, reads may align to both flanking contigs, and provide direct evidence for a "splint" across the gap.

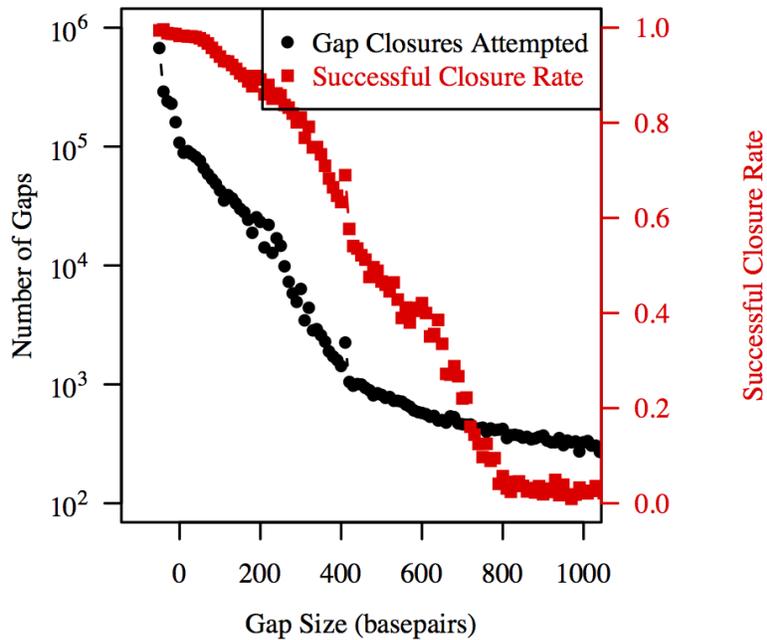

**Figure 5. Gap closures.** Frequency of intra-scaffold gaps before gap closure (green), with gap sizes estimated by a new method. Note that gaps can be negative if the flanking contigs overlap by less than k-1 nucleotide or the k-mers involved at the tips are not unique extensions of each other. Rate of gap closure (red) is also shown as a function of predicted gaps size. The majority of gaps shorter than a read length are closed, but gaps longer than the common fragment size are less frequently closed.

Meraculous makes a series of gap closure attempts for each gap, using increasingly aggressive approaches (Methods). In polymorphic mode, two paths across the gap are allowed; in haploid mode if two paths are identified, the gap remains unclosed. The rate of successful closure is high for negative gaps and gaps shorter than the insert size, and decreases with increasing size, although some long gaps are closed (**Figure 5**).  The eventual contiguity of the assembly is determined by the aggressiveness of gap closure, and to some extent local base-pair accuracy can be exchanged for bulk contig "N50" statistics.

**Speed and memory usage**.  We assembled the human NA12878 genome in one day of "wall clock" time on a compute cluster where resources were requested dynamically.  The largest parallel job set, merblast, consisted of 288 16-way threaded tasks, each requiring up to 17.5 Gb of memory.   The most memory-intensive job, UUcontig determination, was a single 16-way threaded task requiring 111 Gb.   All threaded tasks were run on nodes with 16 cores and 128GB memory, while non-threaded tasks ran on 8-core, 48GB memory nodes.  All stages other than UU contig generation are embarrassingly parallel and can be run as a collection of independent jobs distributed across arbitrary numbers of low memory nodes.  For CPU-intensive operations, most stages include multithreaded code to take full advantage of the compute cycles on each node.  Each node is typically assigned a subset of reads (or k-mers) in a load-balanced manner to manage the memory load.[16] Distributing theses calculations across a larger cluster achieves linear speedup. Details of timings and memory usage are provided in **Table 1**.

**Table 1: Resources used to perform human assembly on Genepool cluster.**

| assembly pipeline stage | Maximum number of cores used in parallel | Maximum single-node memory (GB) | Total wall time (hours) | Total cpu time (core-hours) |
|---|---|---|---|---|
| import | 16 | 2.0 | 2.13 | 3.32 |
| mercount | 256 | 19.0 | 1.85 | 24.75 |
| mergraph | 256 | 16.1 | 2.47 | 31.04 |
| ufx | 16 | < 0.1 | 1.49 | 15.55 |
| contigs | 16 | 106.2 | 4.03 | 4.03 |
| bubble | 256 | 15.4 | 0.48 | 3.47 |
| merblast | 4608 | 16.7 | 3.65 | 243.00 |
| ono | 36 | 47.0 | 4.24 | 24.45 |
| gap_closure | 581 | 14.1 | 3.44 | 566.09 |
| TOTAL | -- | -- | 23.78 | 91 |

Each job can be run on a single node, up to 16 threads/node. For some stages the code is multithreaded and a single 16-way threaded job runs on a single node. Total wall time excludes cluster scheduling wait time.

**Adaptation to AWS/SMP.** Originally designed for running on a large (>1000 node) distributed cluster, Meraculous has now been adapted to a single-machine Symmetric Multiprocessing (SMP) setup within the Amazon Web Services (AWS) environment **(Supplementary Note 5)**. Our goal is to increase the accessibility of Meraculous by leveraging the worldwide availability and standardized environment of the AWS platform. This version consolidates several stages to reduce I/O, uses a third-party memory allocator and pre-sized hash tables to minimize locking at the memory allocation level, and fits the shape of its parallelized work to a single machine to reduce idle cycles.

For maximum throughput we employ a 4x8 Tb RAID-0 cluster configured with the st1 filesystem on a r3.8xlarge instance with 32 cores and 240 Gb RAM. Benchmarking with CloudWatch confirms that Meraculous approaches the theoretical network read limit of 10 Gigabit/s during read-intensive stages. Over 5 runs on the *H. sapiens* read set, the total running time averaged 24.66 hours with a cost of $65.60 for computation and $43.60 for storage (January 2017). Prices reflect on-demand usage in the us-east-1a region---spot requests, reserved requests, and smaller disks can be used to further reduce costs with no changes to the code.

**Summary and evaluation of human assembly.** Bulk "N50" statistics for human assemblies are shown in **Table 2** for all scaffolds larger than 1 kbp. For comparison, we show the corresponding statistics for the AllPathsLG and SGA assemblies for NA12878 datasets, as published by their authors. We note, however, that the AllPathsLG data requirements are not met for some NA12878 datasets, so that only the original AllPathsLG dataset was used,[9] and not short-insert data later reported along with the SGA assembler.[10] The SGA short-insert-only assembly used only half of the reads from the ~300 bp insert libraries they reported, both for speed reasons and because it was assumed that the depth of coverage from half the reads is already sufficient. In contrast, the Merac-all assembly uses both the original AllPathsLG mate-pair dataset and the fragment-pair dataset reported in the SGA assembly paper, and similarly the Merac_frag assembly uses the complete short insert data reported in the SGA paper. Thus the input datasets are not identical, and the performance comparison is intended to contrast features of the various assemblers, rather than as a direct head-to-head competition. L50 numbers are computed with respect to assembly size (in scaffolds larger than 1 kbp).

**Table 2: Raw statistics, human assemblies**.

| Assembly | Merac_all | APLG | Merac_frag | SGA |
|---|---|---|---|---|
| Total contig length (Mbp) | 2,703.8 | 2,614.9 | 2,672.5 | 2,639.5 |
| Contig L50 (kbp) | 51.7 | 23.9 | 28.6 | 10.5 |
| Total scaffold length (Mbp) | 2,840.0 | 2,786.3 | 2,678.9 | 2,655.5 |
| Scaffold L50 | 9.5 Mbp | 12.1 Mbp | 35.5 kbp | 26.5 kbp |

**Completeness**. To evaluate the completeness of our assembly in the context of the NA12878 haplotype-resolved reference,[17] we focused on loci defined by all distinct 101-mers that appear in both reference haplotypes and in all four assemblies. The maternal reference is 3.04 Gbp long, including 200.6 Mbp in gaps, and contains 2,835,530,888 101-mer loci. (A k-mer locus does not include any Ns). Of these, 2,778,837,998 are distinct as 101-mers, but are not necessarily single-copy. Of these distinct maternal 101-mers, 2,566,963,190 (92.4%) are found in equal frequency in both reference haplotypes. We focus our completeness analysis on this subset of 101-mers to minimize the effect of differential haplotype representation on the assessment. We consider 101-mers that are present in *n* copies in both haplotypes, and report the percentages a:b:c:d of these 101-mers that are (a) missing in the respective assembly; (b) present, but in fewer than *n* copies; (c) present exactly *n* times; (d) present in more than *n* copies in the assembly. All four assemblies capture 92-95% of the single copy 101-mers, with Meraculous surpassing the comparator assemblies from SGA and APLG for short-insert-only and multiple insert sizes, respectively. All four assemblies do a poor job of fully capturing two-or-higher copy sequences, which are either missing entirely (48-57% for 101-mers that are two-copy in both references) or underrepresented (34-48% of two-copy 101-mers are found only once in the assemblies). Higher copy 101-mers in the references are more likely to be absent in the assemblies. Given the uncertainty in copy number in the references, only general trends are likely to be meaningful. Complete results are presented in **Table 3**.

**Table 3. Assessment of 101-mers that are present *n* times in both maternal and paternal haplotypes of NA12878.**

| n | Total 101-mers | Merac_all | | | | APLG | | | | Merac_frag | | | | SGA | | | |
|---|---|---|---|---|---|---|---|---|---|---|---|---|---|---|---|---|---|
| | | 0 | <n | =n | >n | 0 | <n | =n | >n | 0 | <n | =n | >n | 0 | <n | =n | >n |

| 1 | 2,542,159,051 | | 4.7 | - | **95** | 0.01 | | 8.0 | - | **92** | 0.34 | | 5.6 | - | **94** | 0.01 | | 7.2 | - | **93** | 0.001 |
|---|---|---|---|---|---|---|---|---|---|---|---|---|---|---|---|---|---|---|---|---|---|
| 2 | 17,105,910 | | 48 | 48 | **4.3** | 0.13 | | 60 | 34 | **5.9** | 0.27 | | 54 | 43 | **2.9** | 0.08 | | 57 | 42 | **0.77** | 0.004 |
| 3 | 4,065,074 | | 63 | 35 | **1.7** | 0.17 | | 79 | 20 | **0.7** | 0.08 | | 69 | 30 | **1.1** | 0.11 | | 70 | 30 | **0.09** | 0.001 |
| 4+ | 3,633,155 | | 71 | 27 | **0.93** | 0.22 | | 82 | 18 | **0.22** | 0.06 | | 79 | 21 | **0.56** | 0.13 | | 88 | 12 | **0.02** | 0.002 |

For 101-mers that occur the same number of times *n* in both reference haplotypes, we report the number of copies in each assembly. In each cell, the four numbers specify the percentages of such 101-mers that are: absent in an assembly; present in lower copy than in reference; present in exactly the reference copy number (shown in bold); and present in higher copy than in reference.

**Base-level accuracy**. The completeness metrics in **Table 3** measure the fraction of the reference genome that is accurately represented by the assembly. Completeness is reduced both by unrepresented sequence (gaps) as well as by inaccurately represented sequence (errors). To determine the relative influence of these two effects we used the same 101-mer framework to measure the single-nucleotide accuracy of the shotgun assemblies. As a baseline measurement we focus on those 101-mers that appear only once in each reference haplotype and consider how well the identity of the 102nd base is predicted in each assembly. We find that in 2,540,241,656 instances (99.92%) the identical base follows a 101-mer in both reference haplotypes (the remaining instances represent polymorphic loci). We assessed the non-polymorphic positions in each shotgun assembly, and for those 101-mers found to be present (subject to the completeness constraints discussed above) the accuracy of the subsequent base is tabulated in **Table 4**.

**Table 4. Assessment of fidelity of locations following uniquely represented reference 101-mers.**

| mismatch rate per 10 kbp | Merac_all | APLG | Merac_frag | SGA |
|---|---|---|---|---|
| $\mu$ (all) | 1.78 | 2.23 | 1.96 | 2.53 |
| $\mu$ (shared) | 0.74 | 1.42 | 0.95 | 1.65 |
| $\mu_M$ (all) | 1.31 | 1.70 | 1.24 | 0.97 |
| $\mu_M$ (shared) | 0.56 | 1.10 | 0.56 | 0.63 |
| $\mu_M$ (shared, fosmid validated) | 0.15 | 0.75 | 0.16 | 0.35 |
| $\mu_U$ (all) | 0.47 | 0.53 | 0.72 | 1.57 |
| $\mu_U$ (shared) | 0.18 | 0.32 | 0.38 | 1.02 |

For each assembly, the observed rates $\mu$, $\mu_M$, and $\mu_U$ (with $\mu = \mu_M + \mu_U$) are reported for all 101-mers ("all") and those 101-mers found in common between all assemblies ("shared"). Additionally, $\mu_M$ is reported for a restricted set of 101-mers that were also validated by a collection of finished fosmid sequences ("fosmid-validated"). Units are events per 10 kbp for all values, and in all cases a lower value is better.

The accuracy is described here by the parameter $\mu$, the rate at which the subsequent base does not match that predicted by (both) reference haplotypes. This parameter can be subdivided into two components: $\mu_M$ is the rate of specified bases that do not match the reference base, and $\mu_U$ is the rate of unspecified bases (Ns) in the subsequent position of the assembly. As such, $\mu_M$ is a measure of correctness, while $\mu_U$ is a measure of local contiguity. While the $\mu$-values calculated for all 101-mers found in each shotgun assembly are directly comparable as rates, due to the completeness effects shown **Table 3** they are normalized by different denominators. We can put all assemblies on an equal footing by restricting ourselves to only those 101-mers that are found in common among all assemblies (there are 2,252,042,044 such 101-mers).

The observed $\mu_M$ rates of ~1 mismatch per 10 kbp for all assemblies are significantly higher than one might expect (for example, as reported in [16] we have observed error rates at least two orders of magnitude lower for Meraculous assemblies with other datasets; see also Assemblathon analyses.[11, 13] Such discrepancies might be expected if the reference haplotypes contained a significant rate of previously-unidentified polymorphism (or even outright error) with respect to individual NA12878. To assess this effect we extended the analysis to consider 103 previously finished fosmid sequences totalling 3,926,558 bp from the individual NA12878. B13. Treating the finished fosmids as we treated the shotgun assemblies, and considering 101-mers found in all assemblies with identical extensions predicted in both parental haplotype references (2,959,240 informative 101-mer loci), the mismatch rate, $\mu_M$, measured for the fosmids sequences is found to be 0.50 per 10 kbp, suggesting that (1) this is the lower bound of accuracy that may be measured relative to the reference haplotypes and that (2) the true accuracy of the shotgun assemblies is significantly higher than what might be estimated here based on the reference haplotype sequences alone. Even these fosmid-validated measurements are likely upper bounds as some of the mismatches may in fact represent polymorphic locations not identified in the fosmids or the reference haplotypes.

**Scaffolding accuracy.** While scaffold N50 is often used to measure the long-range linkage captured in an assembly, this measure conflates true and false linkages. In the worst case scenario, an assembly can have a long scaffold N50 but have many linkage errors, for example, by making indiscriminate joins regardless of supporting data to produce long scaffolds.

A coarse chromosome-scale assessment of scaffold correctness can be performed by simply counting the number of distinct reference chromosomes that share single-copy 101-mers with each scaffold. We expect that for properly assembled scaffolds all of its 101-mers should derive from a single chromosome, except for the occasional stray marker. We use as markers a random sample of 2,253,035 unique 101-mers that are present at single copy in both reference haplotypes and all shotgun assemblies (0.1% of all such 101-mers, with a mean density of 1 per 1.235 kbp in the reference sequence) and consider scaffolds with 10 or more such markers. A scaffold contains a linkage error if it is associated with a significant number of markers (here chosen to be 10 but the analysis is not particularly sensitive to the choice of this threshold) that map to multiple chromosomes. The fraction of total scaffold length that may be assigned to a unique chromosome ($f_U$), to multiple chromosomes ($f_M$) - indicating a gross inter-chromosomal misjoin, or to no chromosome ($f_N$) are reported in **Table 5**. The large misjoin rate for APLG is due to the fact that 24 of the 100 longest scaffolds (which span 63% of the assembly length) are grossly misassembled as measured by 10 or more markers being assigned to different chromosomes (these comprise at least 29 separate misjoins). By comparison, only 3 of the 100 longest scaffolds in the corresponding Meraculous assembly (which span 56% of the assembly length) are grossly misassembled by this measure. (This measure is akin to the optical map concordance metric developed in the Assemblathon II,[11] which also reported high scaffolding accuracy for Meraculous.)  The short-insert only comparison of Meraculous and SGA shows a much lower misjoin error rate (no misjoin errors for Meraculous and only 0.4% of sequence with a significant inter-chromosomal misjoin for SGA, including 21 of the 1000 longest SGA scaffolds). Since these short-insert-only assemblies are generally more fragmented, they also have a higher rate at which scaffolds cannot be assigned to chromosomes by 10 or more markers.

**Table 5.  Assessment of scaffold correctness.**

| scaffolding accuracy measure | Merac_all | APLG | Merac_frag | SGA |
|---|---|---|---|---|
| $f_U$ (%) | 94.4 | 79.2 | 82.9 | 77.5 |
| $f_M$ (%) | 2.7 | 19.3 | 0 | 0.4 |
| $f_N$ (%) | 2.9 | 1.5 | 17.1 | 22.2 |
| $\pi_0$ (per 10,000 markers) | 0.28 | 7.24 | 0.23 | 4.44 |
| $\pi_1$ (per 10,000 markers) | 0.93 | 8.30 | 0.51 | 4.93 |
| $\pi_2$ (per 10,000 markers) | 9.98 | 38.8 | 0.72 | 5.39 |
| $\rho_0$ (per 1,000 markers) | 3.03 | 4.86 | 57.9 | 72.5 |
| $\rho_1$ (per 1,000 markers) | 3.06 | 4.90 | 57.9 | 72.5 |
| $\rho_2$ (per 1,000 markers) | 3.09 | 6.17 | 57.9 | 72.5 |

For each assembly, the fraction of scaffold length that may be assigned to zero ($f_N$), one ($f_U$), or multiple ($f_M$) chromosomes based on the presence of at least 10 shared markers from a sample of 0.1% of all well-localized 101-mers is a "bulk" measure indicating the overall scaffolding correctness and contiguity. The (inverse) scaffold order recall error rates $\rho_0$, $\rho_1$, $\rho_2$ are shown in units of missing associations per 1,000 markers (or, roughly, per 1.2 Mbp). The (inverse) scaffold order precision error rates $\pi_0$, $\pi_1$, $\pi_2$ are shown in units of misorderings per 10,000 markers (or, roughly, per 12 Mbp).

We can also quantify scaffold quality using a recall/precision framework. We first sort the 2,253,035 single-copy 101-mer markers by their order in the reference haplotypes (their order in the two haplotype references is identical). For each marker we then compare the identity of the next marker encountered on an assembled scaffold with its position in the reference genome (precision) or, conversely, consider consecutive markers on the reference and ask how they are represented in the assembly (recall). The scaffold order precision rate can be defined at three levels of precision: $P_0$, the rate at which adjacent scaffold markers are found on the same reference chromosome, $P_1$, the rate at which adjacent scaffold markers are found on the same reference chromosome and the same strand, and $P_2$, the rate at which adjacent scaffold markers are found on the same reference chromosome and the same strand and adjacent to each other. Conversely, we can also define three levels of scaffold order recall: $R_0$, the rate at which adjacent reference markers are found on the same scaffold, $R_1$, the rate at which adjacent reference markers are found on the same scaffold and the same strand, and $R_2$, the rate at which adjacent reference markers are found on the same scaffold and the same strand and adjacent to each other. For each of these quantities we also define an error rate (e.g., $\pi_0 = 1 - P_0$ ; $\rho_0 = 1 - R_0$, *etc.*) for convenience of reporting.

**Table 5** summarizes the results of these assessments. Scaffold order precision error measures the rate at which misjoins are made per 10,000 markers (roughly 12 Mbp). Consistent with the $f_M$ metric, APLG's misjoin rates are high compared with Meraculous using long-insert datasets. While misjoin rates are lower for both Meraculous and SGA with short-insert data than the two long-insert assemblies, Meraculous misjoin rates are an order of magnitude lower than those of SGA. Scaffold recall error rates (which roughly speaking measure the rates at which adjacent markers in the reference are <u>not</u> found in the assembly due to scaffold termination or misjoins) are comparable but a factor of two lower for Meraculous than APLG on the long-insert dataset. Since scaffolds are shorter for short-insert-only asemblies, the recall error rate is substantially higher, but marginally better for Meraculous due to its longer (error-free) scaffolds .

**DISCUSSION**

Genome assemblers aim to reconstruct a complete and accurate genome sequence from a collection of redundant short sequence fragments, using manageable computational resources. Although assembly is algorithmically difficult in a formal sense,[20] following on the pioneering work at TIGR{Sutton [21] 1995} and Celera,[22] and invigorated by the introduction of deBruijn-graph-based approaches[4, 5] there are now numerous practical algorithms that produce broadly useful genome assemblies from data generated using the current generation of short-read sequencers, and short read assembly continues to be an active area of research.[11, 13] Despite this progress, there is room for improvement of contiguity, accuracy, speed, and required computing resources.

Here we described an improved version of our Meraculous assembler[16] that can assemble large and polymorphic genomes with modest computational resources. Meraculous incorporates aspects of both deBruijn and overlap-layout-consensus approaches. The deBruijn approach is used to rapidly identify stretches of the genome whose assembly is uncontested. We do not include an (explicit) error correction step, but rather take advantage of quality scores, depth, and reciprocity to suppress deBruijn graph "hair." Polymorphism is handled by noting that the local sequence around each allele is uncontested, but a locus is represented by a bubble structure with uncontested extensions at either end. This condition allows polymorphism-induced bubbles to participate in uncontested walks through the graph.

We have continued to improve gap-closing performance relative to the original Meraculous, with resulting gains in contig N50 length. Using the human genome as a test set, our results compare favorably with other leading assemblers (AllPathsLG[9] and SGA[10]) that have been run on these same datasets. Due to the differences in data requirements, the input datasets for each of the comparisons are not identical, but include roughly comparable types of data (short-insert-only, and a broad range of fragment and mate-pair data). We used published NA12878 assemblies by the groups that developed APLG and SGA, rather than attempting to rerun or force these assemblers to use exactly the same datasets as Meraculous.

At the start of an assembly project, it is often unclear what kind of assembly the data will support. Here we use the k-mer frequency distribution around single copy depth, and the cumulative frequency distribution over all depths, as a useful diagnostic of what to expect. By fitting the k-mer frequency distribution around the main peak to the sum of two Gaussians at single and half depth, we can compute an estimate of polymorphism rate prior to assembly (**Figure 2b），** which may influence the users choice of subsequent assembly algorithm (**Supplementary Note 2**).

Similarly, the cumulative frequency distribution (**Figure 2c**) provides both an estimate of genome size and the fraction of the genome that is uniquely accessible at the specified value of k. The human genome turns out to be "easy," in the sense that at the modest choice of k=51, 85% of the reference genome is unique; for k=101 this increases to 90%. We can estimate this unique fraction from the cumulative k-mer distribution of random shotgun sequence (**Figure 2b**). For example, the fraction of the genome covered by 51-mers with a count of 1.5x the peak depth or less is 83%, close to the 85% computed directly from the reference. This provides a useful way to infer the expected uniquely-assemblable fraction of a genome simply from k-mer counts. We note however that the uniquely accessible fraction of the genome at a given k is not directly predictive of assembly contiguity, as this also depends on the relative distribution of unique and repetitive sequences along the genome.

One of the challenges of genome assembly is to properly represent the various aspects of assembly quality. The most commonly used bulk metrics are the contig and scaffold "N50" lengths (sometimes "L50") that measure the length of contig or scaffold such that half of the assembled nucleotides are in pieces of this size or longer (and half are in shorter pieces). While these are useful as summary statistics, it is widely recognized among developers and users of assembly algorithms that "N50" numbers are insufficient. In particular, contig N50 can always be increased by "aggressively" closing gaps. In the extreme, one could fill gaps with random sequence, which would inflate contig N50 lengths up to the scaffold N50 size, at the expense of introducing sequence errors. Similarly, one could "aggressively" make poorly supported joins between scaffolds to increase scaffold N50 size, at the expense of creating chimeric sequences with erroneous linkages.

It is therefore essential in evaluating and comparing assemblies that N50 length statistics are also accompanied by some measure of assembly accuracy, to quantify the inherent tradeoffs made by the assembly algorithm. Since for many de novo genome projects few resources are devoted to quantifying accuracy either at the local or global level, users rely on benchmarks, and an assumption that accuracies measured in head-to-head comparisons like the Assemblathons[11, 13] will be representative and can guide users. Yet there is no strong consensus on which measures of accuracy should be used to evaluate assemblies.

Here we presented a methodology for assessing assembly quality compared with a (polymorphic) reference that does not rely on explicit alignments, but rather uses markers to identify common sequences across assemblies and references to provide direct comparison at representative loci. We use 101-mers for these markers, but any odd word length would be sufficient. By measuring the fraction of 101-mers found in each assembly, we can derive a measure of completeness, and assess the degree to which multicopy sequences are missed in the assembly. By measuring the agreement of the bases flanking 101-mer markers found in common between assemblies and references, we can assess base level accuracy. And by observing the rate at which pairs of nearby or distant markers are found together in assemblies, we can measure local and long-range scaffolding accuracy. Since the marker length is longer than the k-mer size used in assembly (51 for the human assembly reported here) the marker provides useful additional information about completeness and accuracy that is not expected to be biased as it would be if the same k were used for the assembly and validation.

While Meraculous produces comparable scaffold "N50" numbers relative to other best-in-class assemblers like AllPathsLG, it has far fewer long-range misjoins, so these long scaffolds are more likely to correctly reflect linkage. Nevertheless, meraculous assemblies are not free of such errors. Manual inspection of these rare errors (~1 gross misjoin per 40 megabases) indicates that these misjoins often involve UU contigs that represent low-but-not-single-copy sequences in the genome.  These UU contigs should be placed in multiple locations in the assembly, but are only placed in a single location, leading to chimeric scaffolds.  (These mis-assemblies could be detected (and broken) now by post-assembly filters that confirm paired-end coverage across each position in the assembly, but this would simply sweep the algorithmic issue under the rug.)  More attention to handling such low copy sequences in their appropriate locations will not only improve scaffolding accuracy further, but will also improve the representation of multi-copy sequences in Meraculous assemblies.

At the contig level, there are "negative" gaps that do not close despite substantial predicted overlap of the flanking sequences, typically because these flanking sequences do not match identically.  Inspection shows that many of these cases are flanked by long homopolymer runs (typically polyA or polyT) that are known to be indel-error prone in Illumina sequencing.[23]  (As with identifying misjoins, we could attempt to fill unclosed gaps by a post-assembly step of read alignment and consensus making).  An improved error model would allow us to recognize these likely errors and take corrective action within the assembly algorithm proper.  Detailed analysis of residual gaps also finds that some are in high GC regions that are known to be difficult for Illumina reads to cover, so some of these gaps are unclosed because of data limitations.

Several developments are ongoing to improve computing performance.  As described here, Meraculous can assemble a human genome using only a modest amount of computing resources, requiring just over 100 GB of peak single-node memory, and more modest memory for the great majority of the computation, rather than the larger shared memory that is required by some other assemblers. Thus it should be portable to most small bioinformatics clusters. Nevertheless, the computing requirements can be pushed down further. In our current "embarrassingly parallel" implementation of the algorithm, each node must read the entire input dataset. This allows each node to compute independent of the other nodes, and in fact the computation can be done serially on a single node, but with many nodes bandwidth may become limiting. A more efficient approach is to read the data in once, and distribute it across multiple nodes, as well as parallelizing the contig traversal step across nodes, which allows the calculation to be distributed across an arbitrary number of small-memory nodes, with near perfect scaling.[24] The mapping of reads back to contigs in preparation for scaffolding can also be distributed in this fashion with perfect scaling.[25] The parallelization of the remaining steps of meraculous for a high performance computing environment is ongoing, and we anticipate that strong scaling over thousands of cores will enable rapid assemblies of human or larger genomes.

**Software implementation and availability**

Meraculous2 software is available in a Perl & C++ implementation. Assemblies are driven via a pipeline engine that utilizes a highly transparent and portable model of job control and monitoring where different assembly stages can be executed and re-executed separately or in unison. The software supports "local" and "cluster" modes of execution where the latter can be configured to run on any Grid Engine-line cluster with relatively little effort.

The software, along with installer, user manual, and a test dataset, is freely available at http://sourceforge.net/projects/meraculous20/

For more information, also visit http://jgi.doe.gov/data-and-tools/meraculous/

**Acknowledgements**

We thank Rob Egan, Jeremy Schmutz, Nicholas Putnam, Therese Mitros, Jessen Bredeson, and Scott Rosen for helpful discussions, and Scott Rosen for assistance with figures. Work performed by the US Department of Energy Joint Genome Institute is supported by the Office of Science of the US Department of Energy under contract number DE–AC02- 05CH11231.

## References

1. Miller, J.R., S. Koren, and G. Sutton, *Assembly algorithms for next-generation sequencing data.* Genomics, 2010. **95**(6): p. 315-27.

2. Schatz, M.C., A.L. Delcher, and S.L. Salzberg, *Assembly of large genomes using second-generation sequencing.* Genome Res, 2010. **20**(9): p. 1165-73.

3. Henson, J., G. Tischler, and Z. Ning, *Next-generation sequencing and large genome assemblies.* Pharmacogenomics, 2012. **13**(8): p. 901-15.

4. Chaisson, M.J. and P.A. Pevzner, *Short read fragment assembly of bacterial genomes.* Genome Res, 2008. **18**(2): p. 324-30.

5. Zerbino, D.R. and E. Birney, *Velvet: algorithms for de novo short read assembly using de Bruijn graphs.* Genome Res, 2008. **18**(5): p. 821-9.

6. Simpson, J.T., et al., *ABySS: a parallel assembler for short read sequence data.* Genome Res, 2009. **19**(6): p. 1117-23.

7. Li, R., et al., *De novo assembly of human genomes with massively parallel short read sequencing.* Genome Res, 2010. **20**(2): p. 265-72.

8. Li, Y., et al., *State of the art de novo assembly of human genomes from massively parallel sequencing data.* Hum Genomics, 2010. **4**(4): p. 271-7.

9. Gnerre, S., et al., *High-quality draft assemblies of mammalian genomes from massively parallel sequence data.* Proc Natl Acad Sci U S A, 2011. **108**(4): p. 1513-8.

10. Simpson, J.T. and R. Durbin, *Efficient de novo assembly of large genomes using compressed data structures.* Genome Res, 2012. **22**(3): p. 549-56.

11. Bradnam, K.R., et al., *Assemblathon 2: evaluating de novo methods of genome assembly in three vertebrate species.* Gigascience, 2013. **2**(1): p. 10.

12. Genomes Project, C., et al., *A map of human genome variation from population-scale*

*sequencing.* Nature, 2010. **467**(7319): p. 1061-73.

13. Earl, D., et al., *Assemblathon 1: a competitive assessment of de novo short read assembly methods.* Genome Res, 2011. **21**(12): p. 2224-41.

14. Salzberg, S.L., et al., *GAGE: A critical evaluation of genome assemblies and assembly algorithms.* Genome Res, 2012. **22**(3): p. 557-67.

15. Genome, K.C.o.S., *Genome 10K: a proposal to obtain whole-genome sequence for 10,000 vertebrate species.* J Hered, 2009. **100**(6): p. 659-74.

16. Chapman, J.A., et al., *Meraculous: de novo genome assembly with short paired-end reads.* PLoS One, 2011. **6**(8): p. e23501.

17. Rozowsky, J., et al., *AlleleSeq: analysis of allele-specific expression and binding in a network framework.* Mol Syst Biol, 2011. **7**: p. 522.

18. Haraksingh, R.R., et al., *Genome-wide mapping of copy number variation in humans: comparative analysis of high resolution array platforms.* PLoS One, 2011. **6**(11): p. e27859.

19. Duitama, J., et al., *Fosmid-based whole genome haplotyping of a HapMap trio child: evaluation of Single Individual Haplotyping techniques.* Nucleic Acids Res, 2012. **40**(5): p. 2041-53.

20. Myers, E.W., *Toward simplifying and accurately formulating fragment assembly.* J Comput Biol, 1995. **2**(2): p. 275-90.

21. Sutton, G.G., et al., *TIGR Assembler: A new tool for assembling large shotgun sequencing projects.* Genome Science and Technology, 1995. **1**(1): p. 9-19.

22. Myers, E.W., et al., *A whole-genome assembly of Drosophila.* Science, 2000. **287**(5461): p. 2196-204.

23. Albers, C.A., et al., *Dindel: accurate indel calls from short-read data.* Genome Res, 2011. **21**(6): p. 961-73.

24. Georganas, E., et al. *Parallel de bruijn graph construction and traversal for de novo*

48genome assembly. in *Proceedings of the International Conference for High Performance Computing, Networking, Storage and Analysis (SC'14)*. 2014.

25. Georganas E., Buluç A., Chapman J. et al. *HipMer: an extreme-scale de novo genome assembler* in *27th ACM/IEEE International Conference on High Performance Computing, Networking, Storage and Analysis, Austin, TX, USA, November 2015.*

**Supplementary Notes for Chapman et al: "Meraculous 2: fast accurate short-read assembly of large polymorphic genomes"**

**Supplementary Note 1. Estimating genome size and heterozygosity**

**Genome size**. If we knew the copy number n of each k-mer in a genome, we could compute the total (haploid) genome size G as

$$G = \sum_n n \times (No.\, of\, k-mers\, with\, copy\, count\, n)$$.

In practice, we do not know the copy count in the genome but can measure the k-mer frequency distribution $f_k(m)$ in a shotgun dataset, i.e., the number of k-mers that occur exactly m times in the dataset. The estimated genomic copy number of a k-mer is $n \sim m/d_0(k)$ where $d_0(k)$ is the k-mer frequency corresponding to a k-mer that is found in a single copy in the genome. Note that the k-dependent single copy k-mer frequency $d_0(k)$ is approximately equal to $N(L-k+1)/G$ where $N$ is the number of reads and $L$ is the read length, but in practice is somewhat lower due to sequencing errors. We estimate $d_0(k)$ from the peak in the k-mer frequency distribution.

We can then estimate the genome size from the k-mer frequency distribution as

$$G \doteq \sum_m \frac{m}{d_0(k)} \times f_k(m).$$

In practice, we must exclude from the sum the *k*-mers that occur at low frequency, since these do not represent *k*-mers present in the genome but arise instead from sequencing errors. As a practical matter we cut off the sum below $m_{min} \sim d_0(k)/4$. Note that for different values of *k*, we will have a different frequency distribution $f_k(m)$, but expect that the estimated genome size will be independent of choice of *k*.

For the human genome, using short-insert data alone, we arrive at the following estimates of genome size by empirically summing the weighted frequency distribution, and also estimate the fraction of the genome that is "k-mer unique" by performing the summation from $m_{min} \sim d_0(k)/4$ up to $1.5 \times d_0(k)$:

**Table S1. Genome size estimates from k-mer counts**

| k  | G (Gbp) | Fraction of genome in single copy k-mers |
|----|---------|------------------------------------------|
| 41 | 2.95    | 86.5%                                    |
| 51 | 2.92    | 88.9%                                    |
| 61 | 2.99    | 90.1%                                    |

We note that many k-mers that appear in two or more copies in the genome can be assembled using meraculous if they occur in a unique genomic context, that is, are surrounded by single-copy k-mers so that they can be accessed by gap closure.

**Heterozygosity**. To estimate the heterozygosity of a shotgun sample of a diploid genome, we note that k-mers overlapping homozygous positions are "full depth" (denoted here $d_0(k)$) but k-mers overlapping heterozygous positions have "half depth" ($d_0/2$). Each isolated heterozygous single nucleotide variant removes one *k*-mer from the full depth peak and adds $2k$ distinct *k*-mers to the half-depth peak, since the polymorphic nucleotide can occur at *k* different locations within the *k*-mer and each allele contributes; a short indel of length $L_{indel}$ adds $2k+L_{indel}$ such *k*-mers. For each k, we fit the distribution as a weighted sum of two normal distributions centered at $d_0$ and $d_0/2$, with a peak-specific width that models the non-uniformity of genome sampling:

$$f(x) = a_1 e^{-\frac{1}{2}\left[\frac{x-d_0}{b_1}\right]^2} + a_2 e^{-\frac{1}{2}\left[\frac{x-\frac{1}{2}d_0}{b_2}\right]^2}.$$

The fit extends over the region around full and half depth, excluding the low depth peak at zero, which represents sequencing error, and extending out to about 1.5x the peak depth. For $k=41$ and the SGA short-insert dataset, the fit parameters were (over the range [10:70], showing standard error of each estimate)

$a_1 = 0.154 \pm 0.000342 \, (0.22\%)$

$a_2 = 0.0316 \pm 0.000444 \, (1.40\%)$

$b_1 = 11.752 \pm 0.0363 \, (0.31\%)$

$b_2 = 9.639 \pm 0.174 \, (1.80\%)$

$d_0 = 47.992 \pm 0.0395 \, (0.082\%)$

For $k=51$ over the range [10:55]

$a_1 = 0.175 \pm 0.000653 \, (0.37\%)$

$a_2 = 0.0448 \pm 0.000757 \, (1.69\%)$

$b_1 = 10.292 \pm 0.0475 \, (0.46\%)$

$b_2 = 8.930 \pm 0.213 \, (2.39\%)$

$d_0 = 39.467 \pm 0.0501 \, (0.127\%)$

For $k=61$ over the range [5:45]

$$a_1 = 0.201 \pm 0.000896 \; (0.44\%)$$
$$a_2 = 0.0612 \pm 0.00150 \; (2.45\%)$$
$$b_1 = 8.903 \pm 0.0628 \; (0.70\%)$$
$$b_2 = 7.922 \pm 0.164 \; (2.07\%)$$
$$d_0 = 31.162 \pm 0.0582 \; (0.19\%)$$

Under a simple infinite alleles coalescent model with effective population size $N$ and population scaled per site mutation rate $\theta = 4N\mu$, the probability that a k-mer locus on the genome is heterozygous (and therefore contributes two counts to the half-depth peak) is $\theta k / (1 + \theta k)$ and the probability that the locus is homozygous (and therefore contributes one count to the full-depth peak) is $1/(1+\theta k)$. So the weighted ratio of the half depth to full depth peak $a_2 b_2 / a_1 b_1$ is then an estimate of $2\theta k$. In this way we estimate $\theta \approx 0.002$. This is about 50% higher than the usually accepted scaled mutation rate for humans, but the k-mer based estimate also includes the effects of indels, which are weighted more than single nucleotide polymorphisms in this scheme.

## Supplementary Note 2. Gap size estimation.

Here we present an unbiased estimate of the size of a gap $g$, for a set of spanning pairs drawn from a known insert size distribution $P(x)$. The two flanking contigs are of finite length $c_1$ and $c_2$ and without loss of generality we assume $c_1 \leq c_2$. The paired-end reads are assumed to have fixed length $L$, and a k-base pair alignment is required between each read and its associated contig.

The mean size of gap-spanning pairs (from which the gap size may be directly estimated) is

$$\langle L_{spanning} \rangle = \frac{\sum x h(x) P(x)}{\sum h(x) P(x)},$$

where the summations (or integrals) are performed over (observable) values of $x$, the end-to-end length of a read pair, and $h(x)$ is the "density of states" defined as:

$$h(x) = \begin{cases} 0: & x \leq x_1 \\ x - x_1: & x_1 \leq x \leq x_2 \\ x_2 - x_1: & x_2 \leq x \leq x_3 \\ x_4 - x: & x_3 \leq x \leq x_4 \\ 0: & x \geq x_4 \end{cases}$$

where

$$x_1 = g + 2k - 1$$
$$x_2 = g + C_1 + L$$
$$x_3 = g + C_2 + L$$
$$x_4 = g + C_1 + C_2 + 2L - 2k + 1$$

The "density of states" $h(x)$ described in the original Meraculous publication is a special case of this, where $x$ is assumed to always be less than $x_2$ (i.e., $C_1$ and $C_2$ are effectively infinite). In this model, we include only the effects of the finite site of the flanking contigs. A "complete" theory (for gaps within finite scaffolds) would allow $h(x)$ to have an arbitrary set of step-function dropouts in gap regions defined by the internal structure of the scaffolds.

If the insert size distribution $P(x)$ is taken to be the normal (Gaussian) distribution with mean $m$ and standard deviation $s$ the summation above may be formally integrated as follows:

$$\langle L_{spanning} \rangle = \frac{N_1 + N_2 + N_3}{D_1 + D_2 + D_3},$$

where

$$N_1 = G_2(x_1, x_2) - x_1 G_1(x_1, x_2)$$
$$N_2 = (x_2 - x_1) G_1(x_2, x_3)$$
$$N_3 = x_4 G_1(x_3, x_4) - G_2(x_3, x_4)$$

and

$$D_1 = G_1(x_1, x_2) - x_1 G_0(x_1, x_2)$$
$$D_2 = (x_2 - x_1) G_0(x_2, x_3)$$
$$D_3 = x_4 G_0(x_3, x_4) - G_1(x_3, x_4)$$

and

$$G_0(a, b) = s \sqrt{\frac{\pi}{2}} \left\{ erf\left[\frac{b-m}{s\sqrt{2}}\right] - erf\left[\frac{a-m}{s\sqrt{2}}\right] \right\}$$

$$G_1(a,b) = s^2 \left\{ e^{-\frac{1}{2}\left[\frac{a-m}{s}\right]^2} - e^{-\frac{1}{2}\left[\frac{b-m}{s}\right]^2} \right\} + mG_0(a,b)$$

$$G_2(a,b) = s^2 G_0(a,b) + mG_1(a,b) + s^2 \left\{ ae^{-\frac{1}{2}\left[\frac{a-m}{s}\right]^2} - be^{-\frac{1}{2}\left[\frac{b-m}{s}\right]^2} \right\}$$

These equations are solved iteratively and self-consistently for the gap size $g$.

**Supplementary Note 3: Resource management**

**Cluster configuration.** The Meraculous2 pipeline can be run in two modes: local and cluster. The cluster mode utilizes job submission and monitoring scripts that are compatible with Grid Engine-like cluster management software (Univa Grid Engine was used in our development/testing framework). The pipeline's job submission module packages and submits sets of jobs to the cluster as task arrays, pings the scheduler, monitors the job status, and evaluates the return status codes sent back by the scheduler. It re-submits failed tasks n number of times (subject to users setting) and alerts the user of any resource limit overruns, suggesting a course of action. This way the user is effectively insulated from having to interface with the cluster directly. Also, the module and the actual submission scripts it calls are de-coupled from the rest of the pipeline software so as to make it easier for external users to modify them safely or even to replace them altogether.

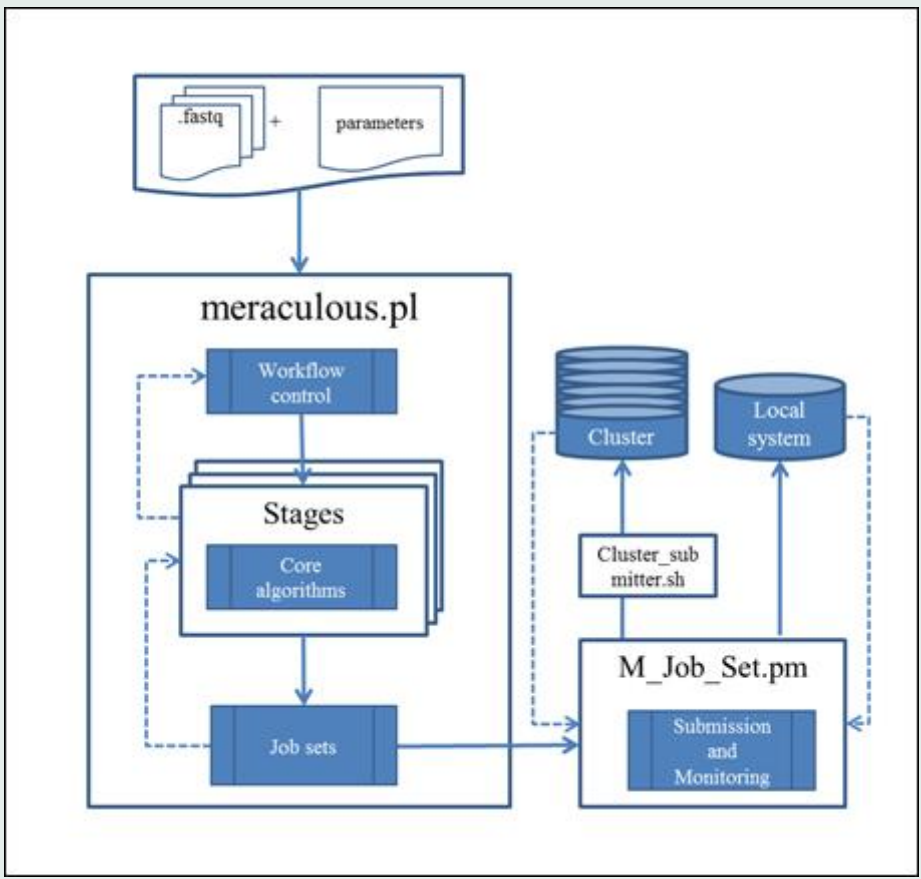

**Fig S1. Pipeline design**

**Resource management.** The Meraculous2 pipeline driver (meraculous.pl) was designed to allow utilization of a wide range of computing architectures, from a modest standalone server to high performance cluster environments. In order for all the processes of the pipeline to fit under a given set of resource limits, as specified by the user, input data is partitioned into blocks of appropriate size early on in the process. The primary target of optimization is memory, since that is assumed to be the limiting resource in most user environments. The current implementation of Meraculous has a single non-parallel component (the UUtig generation) that defines the absolute overall minimum of memory required. All other memory-intensive processes, like k-mer counting, take advantage of data partitioning to ensure that this minimum is not exceeded (Fig. 2).

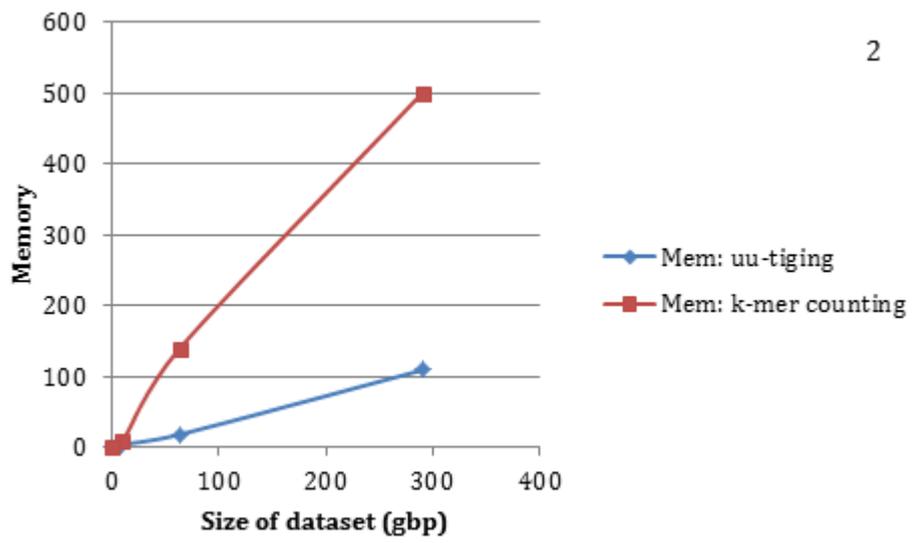

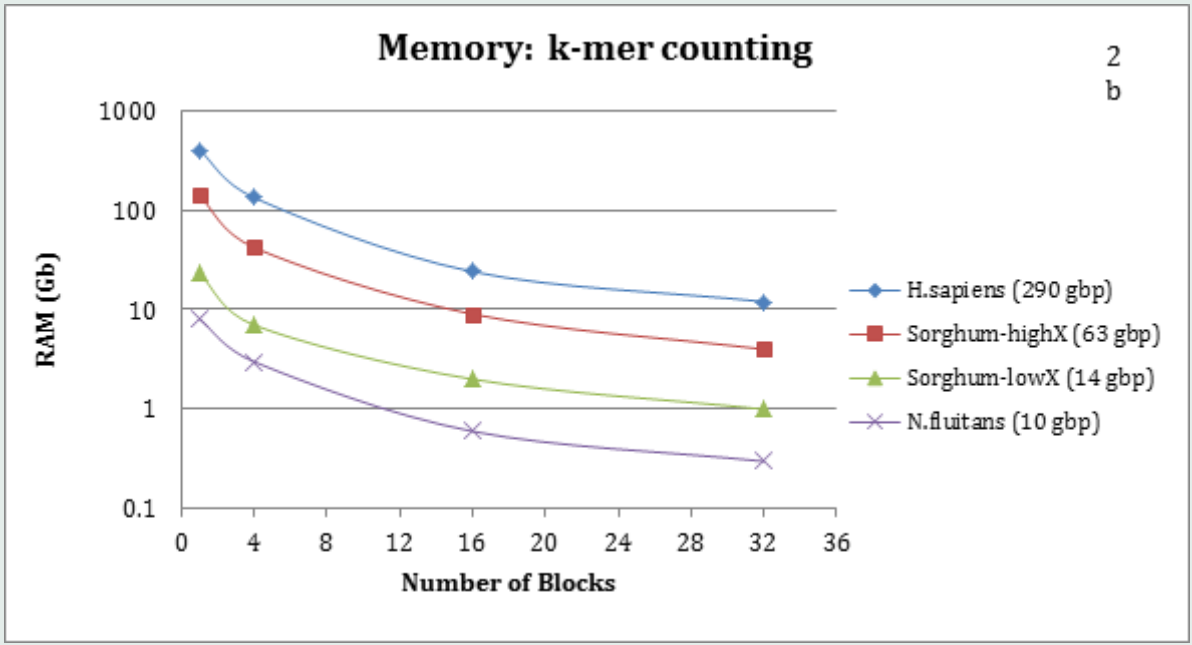

**Fig S2.** Top: memory consumption by non-partitioned k-mer counting and UUtig generation processes. Bottom: memory consumption by k-mer counting in response to partitioning of the k-mer space (notice the log scale of y). Sorghum is a repetitive grass genome (haploid size ~800 Mb; sample was homozygous); *N. fluitans* is a heterozygous fungal genome (haploid size ~60 Mb; sample was highly heterozygous).

Once the data has been partitioned into memory-fitting blocks, the time it takes to complete an assembly is inversely proportional to the number of CPUs available for parallel computation, plus any latency associated with network speed/load, disk IO, or process scheduling.  The data blocks don't need to be processed all at the same time, so, if there aren't enough CPUs available the pipeline engine will simply wait until all the jobs have been scheduled and completed.   This way the user doesn't need to worry about creating too many data blocks – they will all be processed as resources become available.    The benefit of this model is that having low compute resources no longer means 'out of memory' crashes.  It just translates into longer run times.

      The memory footprint is roughly proportional to the size of the input dataset but will vary with sequence quality, read-level duplication, and repetitiveness of the genome.   For instance, the memory required for k-mer counting is 3-4 times higher for the *Sorghum bicolor* "low-X" data set compared to the *Naiadella fluitans* data set of similar value (Table S2).  This can be attributed to the fact that the inbred Sorghum genome is approximately 4 times larger than *N. fuitans*, which leads to a higher number of unique k-mers to store.  The Sorghum data set was also of relatively poor sequence quality, which further inflates the unique k-mer stack.

At this moment the software does not predict the memory requirements up front. Instead we use the number of k-mer blocks as a direct parameter that would be provided by the user. Tables S2 and S3 can be used as a rough guide for estimating the desired number of blocks given an Illumina data set of comparable size.

| Memory: UUtig construction (GB) | | |
|---|---|---|
| H.sapiens | S.bicolor | N.fluitans |
| 111 | 18 | 2.7 |

**Table S1.** Peak memory consumption during the non-parallel UUtig generation step.

| # of blocks | Memory: k-mer counting (GB) | | | |
|---|---|---|---|---|
| | H.sapiens- 290 Gbp | Sorghum bicolor - 63 Gbp | Sorghum bicolor (low-X)- 14 Gbp | Naiadella fluitans - 10 Gbp |

| | | | | | |
|---|---|---|---|---|---|
| 1 | 500 (est.) | 139 | 23 | | 8 |
| 4 | 133 | 42 | 7 | | 2 |
| 16 | 24 | 9 | 2 | | 0.6 |
| 32 | 12 | 4 | 1 | | 0.3 |

**Table S3.** Peak memory consumption during the parallel k-mer counting step.

Many key Meraculous components utilize multi-threading. The user specifies up front how many CPU cores are to be used by threaded programs (if run on a cluster, this means number of cores per node). Again, the pipeline will rely on the cluster software to assign the threaded jobs to nodes with at least (n+1) available cores, while non-threaded jobs will be assigned to single cores. Note that the number of threads selected has no effect on the memory requirement since the size of the data block is determined independently.

The parameters that were used in run_2014-09-15_15h46m21s were as follows:

```
lib_seq  /global/dna/projectdirs/plant/assembly/eugeneg/SEQ_DATA/H.sapiens/used_in
0     1 1 1   0
0
lib_seq  /global/dna/projectdirs/plant/assembly/eugeneg/SEQ_DATA/H.sapiens/used_in
1     0 2 0   0 25
lib_seq  /global/dna/projectdirs/plant/assembly/eugeneg/SEQ_DATA/H.sapiens/used_in
1     0 2 0   0 25
lib_seq  /global/dna/projectdirs/plant/assembly/eugeneg/SEQ_DATA/H.sapiens/used_in
0     0 3 0   45 15
lib_seq  /global/dna/projectdirs/plant/assembly/eugeneg/SEQ_DATA/H.sapiens/used_in
0     0 3 0   45 15

genome_size     3.1
is_diploid      1
bubble_min_depth_cutoff 1
mer_size 51
min_depth_cutoff        4
num_prefix_blocks 16
no_read_validation 1
gap_close_aggressive 1
gap_close_rpt_depth_ratio 3

use_cluster 1
cluster_walltime        12:00:00
cluster_ram_request     24
```

```
cluster_ram_meraculous_import     4
cluster_ram_meraculous_mercount   0
cluster_ram_meraculous_mergraph   0
cluster_ram_meraculous_ufx        2
cluster_ram_meraculous_contigs    128
cluster_ram_meraculous_bubble     0
cluster_ram_meraculous_merblast   0
cluster_ram_meraculous_ono        0
cluster_ram_meraculous_gap_closure    0
cluster_project     plant-assembly.p
cluster_slots_per_task   16
cluster_max_retries      0

local_num_procs          8
local_max_memory         128
local_max_retries                   0
```

**Supplementary Note 4. Summary statistics for assemblies assessed in main text.**

This note summarizes the basic statistics of the four assemblies of NA12878 compared and referenced in the main text. These include: the SGA assembly with short-insert data only (Simpson and Durbin); the comparable meraculous assembly (merac_short) of this short-insert dataset; the APLG assembly with various insert sizes as described in Jaffe et al. (Jaffe et al. ); and the merac_all assembly including both short-insert data and the datasets from Jaffe et al.

| SGA (short insert only) | |
|---|---:|
| Total scaffold sequence | 2,774.3 MB |
| Number of scaffolds | 583,982 |
| scaffold N/L50 | 31,116 / 25.2 KB |
| Total contig sequence | 2,758.1 MB (0.6% gap) |
| Number of contigs | 890,072 |
| Contig N/L50 | 77,657 / 9.9 KB |
| Number of scaffolds > 50 KB | 8,377 |
| % assembly in scaffolds > 50 KB | 21.5% |

| minimum scaffold length | number of scaffolds | number of contigs | total scaffold length | total contig length | scaffold coverage by contigs |
|---|---|---|---|---|---|
| All | 583,982 | 890,072 | 2,774,335,131 | 2,758,094,652 | 99.41% |
| 1 kb | 189,630 | 489,224 | 2,655,532,735 | 2,639,598,976 | 99.40% |
| 2.5 kb | 152,637 | 438,246 | 2,594,792,755 | 2,579,641,834 | 99.42% |
| 5 kb | 120,056 | 383,266 | 2,475,886,276 | 2,461,940,498 | 99.44% |
| 10 kb | 81,123 | 299,598 | 2,193,311,642 | 2,181,768,373 | 99.47% |
| 25 kb | 31,419 | 152,239 | 1,394,789,608 | 1,388,459,523 | 99.55% |
| 50 kb | 8,379 | 53,860 | 596,480,028 | 594,104,454 | 99.60% |
| 100 kb | 819 | 7,622 | 101,002,929 | 100,650,045 | 99.65% |
| 250 kb | 3 | 54 | 820,626 | 818,503 | 99.74% |

| MERAC_short | |
|---|---:|
| Total scaffold sequence | 2,678.9 MB |
| Number of scaffolds | 160,784 |
| Scaffold N/L50 | 21,420 / 35.5 KB |
| Total contig sequence | 2,672.5 MB (0.2% gap) |
| Number of contigs | 210,816 |
| Contig N/L50 | 26,508 / 28.6 KB |
| Number of scaffolds > 50 KB | 11,929 |
| % assembly in scaffolds > 50 KB | 35.1% |

| minimum scaffold length | number of scaffolds | number of contigs | total scaffold length | total contig length | scaffold coverage by contigs |
|---|---|---|---|---|---|
| All | 160,784 | 210,816 | 2,678,919,056 | 2,672,508,920 | 99.76% |
| 100 | 160,784 | 210,816 | 2,678,919,056 | 2,672,508,920 | 99.76% |
| 250 | 160,784 | 210,816 | 2,678,919,056 | 2,672,508,920 | 99.76% |
| 500 | 160,784 | 210,816 | 2,678,919,056 | 2,672,508,920 | 99.76% |
| 1 kb | 160,784 | 210,816 | 2,678,919,056 | 2,672,508,920 | 99.76% |
| 2.5 kb | 124,123 | 166,172 | 2,620,257,356 | 2,615,092,432 | 99.80% |
| 5 kb | 99,684 | 137,276 | 2,531,948,180 | 2,527,435,896 | 99.82% |
| 10 kb | 72,005 | 103,567 | 2,330,247,174 | 2,326,565,597 | 99.84% |
| 25 kb | 33,783 | 53,450 | 1,707,258,802 | 1,705,108,030 | 99.87% |

| APLG (short and long data) | |
|---|---:|
| Total scaffold sequence | 2,786.3 MB |
| Number of scaffolds | 11,393 |
| Scaffold N/L50 | 67 / 12.1 MB |
| Total contig sequence | 2,614.9 MB (6.1% gap) |
| Number of contigs | 231,194 |
| Contig N/L50 | 30,971 / 23.9 KB |
| Number of scaffolds > 50 KB | 744 |
| % assembly in scaffolds > 50 KB | 98.70% |

| minimum scaffold length | number of scaffolds | number of contigs | total scaffold length | total contig length | scaffold coverage by contigs |
|---|---|---|---|---|---|
| All | 11,393 | 231,194 | 2,786,258,565 | 2,614,892,607 | 93.85% |
| 1 kb | 11,389 | 231,190 | 2,786,254,569 | 2,614,888,611 | 93.85% |
| 2.5 kb | 4,260 | 224,057 | 2,776,785,661 | 2,605,420,114 | 93.83% |
| 5 kb | 2,589 | 221,191 | 2,770,456,624 | 2,600,656,012 | 93.87% |
| 10 kb | 1,434 | 218,526 | 2,762,246,777 | 2,593,982,100 | 93.91% |
| 25 kb | 852 | 216,102 | 2,753,960,862 | 2,587,374,736 | 93.95% |
| 50 kb | 744 | 215,275 | 2,749,709,728 | 2,585,141,131 | 94.02% |
| 100 kb | 631 | 213,738 | 2,741,800,729 | 2,580,230,396 | 94.11% |
| 250 kb | 531 | 211,191 | 2,725,851,569 | 2,567,949,642 | 94.21% |
| 500 kb | 444 | 206,887 | 2,695,179,619 | 2,542,359,508 | 94.33% |
| 1 mb | 357 | 198,965 | 2,631,425,091 | 2,486,769,990 | 94.50% |
| 2.5 mb | 250 | 181,959 | 2,456,802,245 | 2,326,826,671 | 94.71% |
| 5 mb | 164 | 154,300 | 2,135,544,572 | 2,027,828,102 | 94.96% |

| MERAC_all | |
|---|---:|
| Total scaffold sequence | 2,840.0 MB |
| Number of scaffolds | 15,873 |
| Scaffold N/L50 | 87 / 9.5 MB |
| Total contig sequence | 2,703.8 MB (4.8% gap) |
| Number of contigs | 149,070 |
| Contig N/L50 | 15,175 / 51.7 KB |
| Number of scaffolds > 50 KB | 957 |
| % assembly in scaffolds > 50 KB | 97.6% |

| minimum scaffold length | number of scaffolds | number of contigs | total scaffold length | total contig length | scaffold coverage by contigs |
|---|---|---|---|---|---|
| All | 15,873 | 149,070 | 2,839,952,726 | 2,703,778,310 | 95.21% |
| 100 | 15,873 | 149,070 | 2,839,952,726 | 2,703,778,310 | 95.21% |
| 250 | 15,873 | 149,070 | 2,839,952,726 | 2,703,778,310 | 95.21% |
| 500 | 15,873 | 149,070 | 2,839,952,726 | 2,703,778,310 | 95.21% |
| 1 kb | 15,873 | 149,070 | 2,839,952,726 | 2,703,778,310 | 95.21% |
| 2.5 kb | 9,714 | 141,016 | 2,831,247,682 | 2,695,371,529 | 95.20% |
| 5 kb | 4,996 | 130,297 | 2,813,646,525 | 2,685,235,703 | 95.44% |
| 10 kb | 2,345 | 121,391 | 2,794,885,372 | 2,673,832,840 | 95.67% |
| 25 kb | 1,155 | 115,445 | 2,778,130,730 | 2,662,903,175 | 95.85% |
| 50 kb | 957 | 113,764 | 2,771,185,691 | 2,658,721,266 | 95.94% |
| 100 kb | 742 | 111,713 | 2,756,222,098 | 2,651,262,737 | 96.19% |
| 250 kb | 591 | 109,046 | 2,732,781,542 | 2,633,384,545 | 96.36% |
| 500 kb | 508 | 106,612 | 2,702,142,491 | 2,606,358,474 | 96.46% |

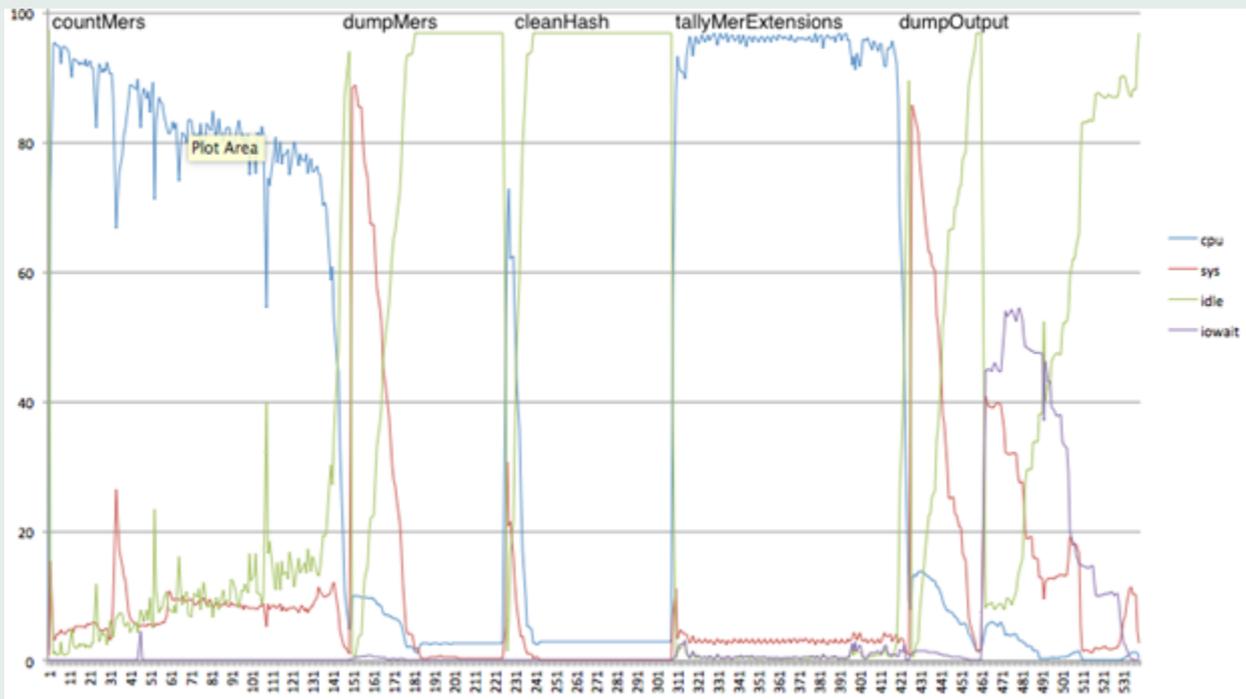

**Fig S3.** Cpu/sys/idle/iowait percentages in mercountUfx on human after throughput optimization. The countMers and tallyMerExtensions correspond to the counting sections of the mercount and mergraph stages of the original pipeline respectively. Throughput optimizations have raised the average cpu percentage to above 80% in countMers and 95% in tallyMerExtensions for a human assembly on r3.8xlarge.

**Supplementary Note 5.  Improving performance on an AWS/SMP architecture**

For AWS, we have developed a separate version of the pipeline to maximize performance on an SMP architecture.  As described previously, several of the stages of Meraculous--in particular the mercount, mergraph and merblast---work extensively across multiple threads on an array of 65536 google sparse_hashes indexed by DNA-prefix to minimize contention. The move to SMP increases contention and I/O load on a per-process basis.  Initial benchmarking of the mercounter stage showed an effective cpu efficiency of merely 40% over the counting portion of a typical mercount and mergraph process. Careful profiling of the iowait / cpu usage statistics, which showed low iowait but heavy idle time despite using spinlocks for access control, revealed that the code was locking-bound at the memory allocation level, particularly when resizing the sparse_hashes during population.

To address this, in the SMP pipeline we a) combine the mercounter, mergraph and UFX stages into a single mercountUfx stage, b) preallocate our hash tables based on known estimates to minimize hash reallocation and c) replace malloc with a third-party memory management library (tcmalloc) designed to perform well under heavy threading.  Note that

sparse_hashes only require 5.3 bits per bin, so an overestimate of 10x the number of compressed 51-mers results in just 20% of memory overhead.

After these changes, throughput in human improved to the point of becoming I/O bound. By using RAID-0 across an array of 4 8Tb st1 volumes, each specified to provide 2.6Gigabit/s of read bandwidth, we were able to effectively saturate the 10 gigabit/s network connection over the counting portion of the mercountUfx stage, with merblast showing markedly improved performance as well (Figure S3).

The previous version of merBlast builds a hash table of all of the kmers in the contigs for each process, with a single process assigned to each node to increase throughput. On a single-machine the inefficiency of rebuilding the hash table becomes dominant, so the SMP version dumps a binary hash at the end of the contig generation stage. Then each merblast process loads in metadata from the bubble stage, restores the hash, then translates coordinates forward to reproduce the post-bubble hash.

On a single machine, the stages post-contig generation require a different parallelization scheme than those upstream of contig generation. While the latter's memory footprint is based on the k-mer hash, the former's is based on the contig set. Using the default parallelization will typically lead to a significant percentage of unused cycles and unneeded multiple batches of jobs to be run. By adding a new num_chunks_ono parameter, we can assign the later stages to a different number of jobs than the upstream stages, thus fitting the single machine case much more efficiently.

To achieve a 50% reduction in memory footprint in the contig generation stage, we hash only the canonical mers, which introduces a small overhead in runtime but allows us to fit the full human data set into a single r3.8xlarge instance on AWS.

Funding for these tunings was provided by an AWS in Education grant.